\documentclass[12pt,preprint]{aastex}
\newcommand{\etal}{et\,al.}

\newcommand{\kms}{km\,s$^{-1}$}
\newcommand{\lsim}{\raise0.3ex\hbox{$<$}\kern-0.75em{\lower0.65ex\hbox{$\sim$}}}

\newcommand{\HI}{H~{\sc I}}
\newcommand{\msun}{M$_{\odot}$}

\usepackage{fancyhdr}
\usepackage{color}
\usepackage[yyyymmdd,hhmmss]{datetime}
\begin{document}
\slugcomment{\color{red}{\textbf{Accepted for Publication in the Astronomical Journal}}}

\title{The ALFALFA ``Almost Darks'' Campaign:  Pilot VLA HI Observations 
of Five High Mass-to-Light Ratio Systems}


\author{John M. Cannon}
\affil{Department of Physics \& Astronomy, Macalester College, 1600
  Grand Avenue, Saint Paul, MN 55105}
\email{jcannon@macalester.edu}

\author{Charlotte P. Martinkus}
\affil{Department of Physics \& Astronomy, Macalester College, 1600
  Grand Avenue, Saint Paul, MN 55105}
\email{cmartink@macalester.edu}

\author{Lukas Leisman}
\affil{Center for Radiophysics and Space
  Research, Space Sciences Building, Cornell University, Ithaca, NY
  14853, USA}
\email{leisman@astro.cornell.edu}

\author{Martha P. Haynes}
\affil{Center for Radiophysics and Space
  Research, Space Sciences Building, Cornell University, Ithaca, NY
  14853, USA}
\email{haynes@astro.cornell.edu}

\author{Elizabeth A. K. Adams}
\affil{Netherlands Institute for Radio Astronomy (ASTRON), Postbus 2,
  7990 AA, Dwingeloo, The Netherlands}
\email{adams@astron.nl}

\author{Riccardo Giovanelli}
\affil{Center for Radiophysics and Space
  Research, Space Sciences Building, Cornell University, Ithaca, NY
  14853, USA}
\email{riccardo@astro.cornell.edu}

\author{Gregory Hallenbeck}
\affil{Center for Radiophysics and Space
  Research, Space Sciences Building, Cornell University, Ithaca, NY
  14853, USA}
\affil{Department of Physics and Astronomy, Union College,
  Schenectady, NY 12308, USA}
\email{hallenbg@union.edu}

\author{Steven Janowiecki}  
\affil{Department of Astronomy, Indiana University, 727 East
  Third Street, Bloomington, IN 47405, USA}
\email{sjanowie@indiana.edu}

\author{Michael Jones}
\affil{Center for Radiophysics and Space
  Research, Space Sciences Building, Cornell University, Ithaca, NY
  14853, USA}
\email{jonesmg@astro.cornell.edu}

\author{Gyula I.G. J{\'o}zsa}
\affil{SKA South Africa, Radio Astronomy Research Group, 3rd Floor,
  The Park, Park Road, Pinelands, 7405, South Africa}
\affil{Rhodes University, Department of Physics and Electronics,
  Rhodes Centre for Radio Astronomy Techniques \& Technologies, PO Box
  94, Grahamstown, 6140, South Africa}
\affil{Argelander-Institut f{\"u}r Astronomie, Auf dem H{\"u}gel 71, 
53121 Bonn, Germany}
\email{jozsa@ska.ac.za}

\author{Rebecca A. Koopmann} 
\affil{Department of Physics and Astronomy, Union College,
  Schenectady, NY 12308, USA}
\email{koopmanr@union.edu}

\author{Nathan Nichols}
\affil{Hartwick College, Oneonta, NY 13820, USA}
\email{nicholsn@hartwick.edu}

\author{Emmanouil Papastergis}
\affil{Kapteyn Astronomical Institute, University of Groningen, 
Postbus 800, 9700 AA, Groningen, The Netherlands}
\email{papastergis@astro.rug.nl}

\author{Katherine L. Rhode}
\affil{Department of Astronomy, Indiana University, 727 East
  Third Street, Bloomington, IN 47405, USA}
\email{rhode@astro.indiana.edu,}

\author{John J. Salzer}  
\affil{Department of Astronomy, Indiana University, 727 East
  Third Street, Bloomington, IN 47405, USA}
\email{slaz@astro.indiana.edu}

\author{Parker Troischt}
\affil{Hartwick College, Oneonta, NY 13820, USA}
\email{troischt@hartwick.edu}

\begin{abstract}

We present new VLA \HI\ spectral line imaging of five sources
discovered by the ALFALFA extragalactic survey.  These targets are
drawn from a larger sample of systems that were not uniquely
identified with optical counterparts during ALFALFA processing, and as
such have unusually high \HI\ mass to light ratios.  The candidate
``Almost Dark'' objects fall into four broad categories: 1) objects
with nearby \HI\ neighbors that are likely of tidal origin; 2) objects
that appear to be part of a system of multiple \HI\ sources, but which
may not be tidal in origin; 3) objects isolated from nearby ALFALFA
\HI\ detections, but located near a gas-poor early-type galaxy; 4)
apparently isolated sources, with no object of coincident redshift
within $\sim$400 kpc.  Roughly 75\% of the 200 objects without
identified counterparts in the $\alpha$.40 database (Haynes et
al. 2011) fall into category 1 (likely tidal), and were not considered
for synthesis follow-up observations. The pilot sample presented here
(AGC\,193953, AGC\,208602, AGC\,208399, AGC\,226178, and AGC\,233638)
contains the first five sources observed as part of a larger effort to
characterize \HI\ sources with no readily identifiable optical
counterpart at single dish resolution (3.5\arcmin). These objects span
a range of \HI\ mass [7.41 $<$ log(M$_{\rm HI}$) $<$ 9.51] and
\HI\ mass to B-band luminosity ratios (3 $<$ M$_{\rm HI}$/L$_{\rm B}$
$<$ 9).  We compare the \HI\ total intensity and velocity fields to
optical imaging drawn from the Sloan Digital Sky Survey and to
ultraviolet imaging drawn from archival GALEX observations.  Four of
the sources with uncertain or no optical counterpart in the ALFALFA
data are identified with low surface brightness optical counterparts
in Sloan Digital Sky Survey imaging when compared with VLA
\HI\ intensity maps, and appear to be galaxies with clear signs of
ordered rotation in the \HI\ velocity fields. Three of these are
detected in far-ultraviolet GALEX images, a likely indication of star
formation within the last few hundred Myrs.  One source (AGC\,208602)
is likely tidal in nature, associated with the NGC\,3370 group.
Consistent with previous efforts, we find no ``dark galaxies" in this
limited sample. However, the present observations do reveal complex
sources with suppressed star formation, highlighting both the
observational difficulties and the necessity of synthesis follow-up
observations to understand these extreme objects.

\end{abstract}						

\keywords{galaxies: evolution --- galaxies: dwarf --- galaxies:
  irregular --- galaxies: individual (AGC193953, AGC208602, AGC208399, AGC226178,
AGC233638)}

\section{Introduction}
\label{S1}

ALFALFA, the Arecibo Legacy Fast ALFA extragalactic \HI\ survey, has
obtained a census of \HI-bearing galaxies over 7000 square degrees of
high Galactic latitude sky visible from Arecibo between $0^\circ <$
$\delta$ $< +36^\circ$ \citep{giovanelli05,haynes11}.  Its combination of
sensitivity, bandwidth and sky coverage make ALFALFA the first wide
area \HI\ survey to sample a cosmologically significant volume,
delivering robust determinations of the \HI\ mass, width, and diameter
functions, the \HI\ correlation function, and the cosmic density of
\HI\ at $z\sim$0 ({Martin \etal\ 2010}\nocite{martin10}; {Dowell
  \etal\ 2011}\nocite{dowell11}; {Papastergis
  \etal\ 2011}\nocite{papastergis11}; {Martin
  \etal\ 2012}\nocite{martin12}).  When completed, the final ALFALFA
database will contain more than 30,000 high S/N \HI\ detections.

As discussed in \citet{haynes11}, the 40\% ALFALFA catalog (hereafter
``$\alpha$.40'') contains 15,855 extragalactic \HI\ sources.  Less
than 1.5\% of these sources were not clearly identified with an
associated stellar population (hereafter referred to as an ``optical
counterpart'') in an optical survey (SDSS or DSS). Of these, only
$\sim$50 were not readily associated with a known structure or nearby
\HI\ source (that is, roughly 75\% of these ``dark'' candidates
are likely tidal in origin). The nature of the remaining
objects remains unclear; their optical associations are often
ambiguous given the ALFALFA pointing accuracy ($\sim$26\arcsec\ for
reasonable S/N sources) and unresolved given the ALFALFA beam
(3.5\arcmin).  As discussed in detail below, there are only a few
viable candidate ``dark'' galaxies in the current literature; the rich
population of sources measured by ALFALFA is a promising place to look
for more.

\begin{figure}
\epsscale{1}
\plotone{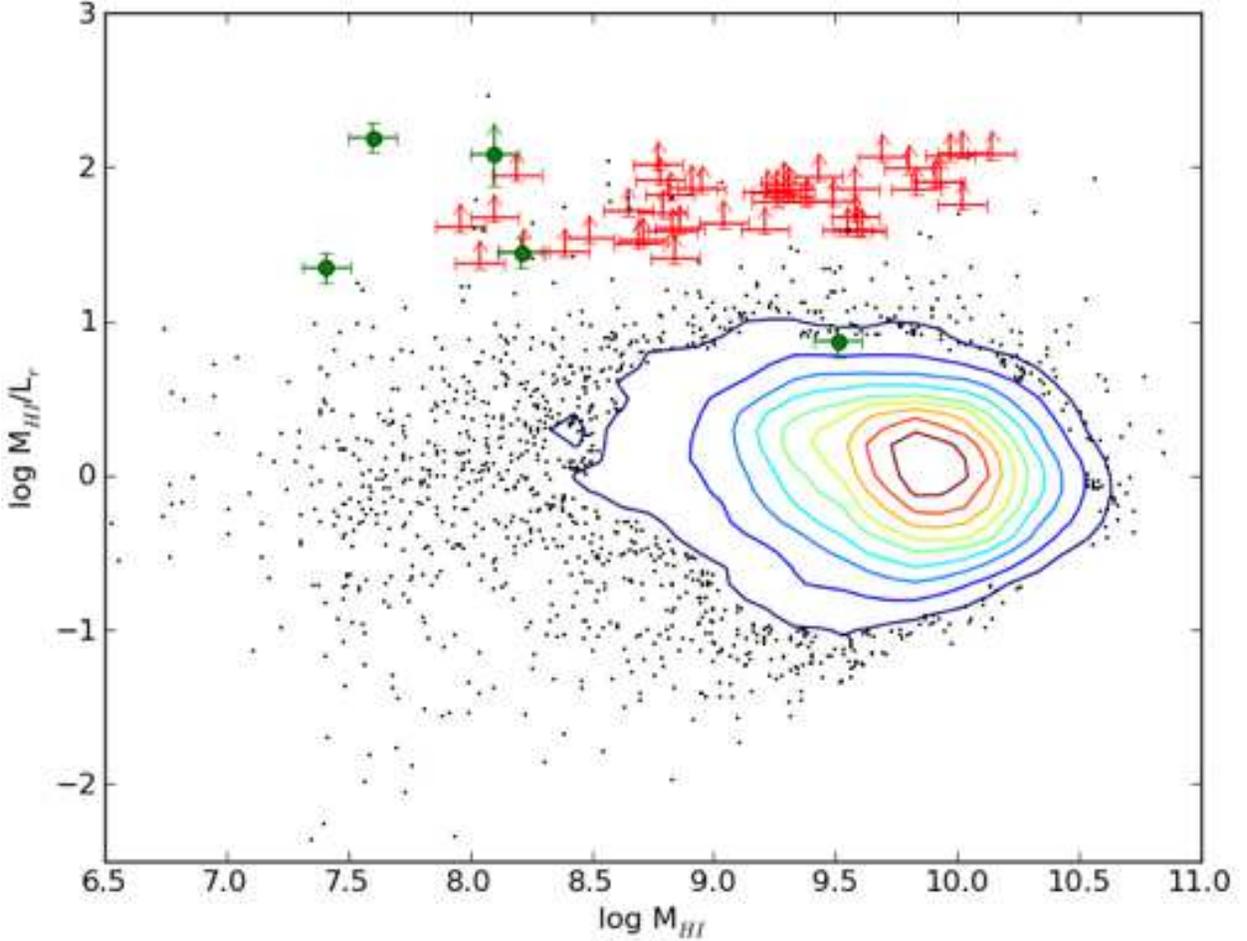}
\caption{Plot of the logarithm of the neutral hydrogen to optical
  luminosity ratio, M$_{\rm HI}$/L$_{\rm B}$, versus the logarithm of
  the \HI\ mass, for all members of the $\alpha$.40 catalog (Haynes
  \etal\ 2011).  All black points and areas within contours represent
  galaxies with reliable SDSS optical counterparts.  The contours show
  number densities in intervals of 10\%, from 10\% to 90\% of the
  number density maximum (which occurs at log(M$_{\rm HI}$) $=$ 9.8;
  see Haynes \etal\ 2011).  The candidate ``Almost Dark'' sources are
  shown in red or green (studied in this work).  These sources have
  high M$_{\rm H}$/L but lack obvious or uncertain optical
  counterparts.  The green point at log(M$_{\rm HI}$) $=$ 9.51 is
  AGC\,233638, a massive \HI\ source with a discrepant SDSS redshift
  whose association was uncertain based on ALFALFA alone; see detailed
  discussion in \S~\ref{S3.3.2}.  For consistency, all M$_{\rm HI}$/L
  values calculated here use SDSS catalog magnitudes.}
\label{images-selection}
\end{figure}

To probe the nature of these candidate ``dark'' objects, we have
initiated the ALFALFA ``Almost Darks'' program: a comprehensive,
multi-wavelength follow-up observing campaign to characterize the
nature of ALFALFA \HI\ detections that were not identified with an
unambiguous optical counterpart in initial ALFALFA processing.  We
anticipate that detailed studies of these objects will result in the
confirmation of associated optical counterparts for the vast majority
of systems.  Figure~\ref{images-selection} shows these objects in
comparison to the full $\alpha$.40 catalog; the 11,300 ALFALFA sources
with good SDSS photometry (see {Haynes \etal\ 2011}\nocite{haynes11})
are plotted as either individual black points, or as colored contours
at 10\% levels (between 10\% and 90\%) of the number density of
sources. Optical luminosities for these sources are calculated using
SDSS catalog magnitudes from the ALFALFA-SDSS cross match presented in
\citet{haynes11}. The 52 candidate ``dark'' sources are shown in red,
with the 5 objects selected for the pilot sample studied in detail in
the present manuscript shown in green. These 52 sources are those that
have no clear optical counterpart assigned in the initial ALFALFA
processing, and that were not deemed as likely to be tidal in
nature. Red points represent upper limits in luminosity estimated at
the limiting magnitude for a source 10\arcsec\ in diameter in
downloaded SDSS images. Green points that are not upper limits are
calculated using the SDSS catalog magnitudes of the optical
counterparts identified after VLA observations; prior to source
localization with the VLA these objects were upper limits similar to
the points plotted in red.

While Figure~\ref{images-selection} was not used for the selection of
candidate ``Almost Dark'' sources, it does clearly demonstrate that
these systems harbor high \HI\ mass to stellar light ratios
(hereafter, M$_{\rm HI}$/L) compared to the vast majority of
$\alpha$.40 detections.  Note that some sources with elevated 
M$_{\rm HI}$/L ratios are not classified as candidate ``Almost Dark''
systems: these sources have clearly associated optical counterparts.
Further, while no strict minimum value of M$_{\rm HI}$/L was enforced
to create the sample of ``Almost Dark'' systems, an empirical
threshold (which does have significant uncertainty at low luminosity
and low surface brightness) appears around M$_{\rm HI}$/L$\simeq$10.

In this manuscript we present pilot VLA \HI\ spectral line
observations of 5 candidate ``Almost Dark'' sources.  This represents
the first step in a major program to understand the nature and origin
of the most extreme M$_{\rm HI}$/L systems within the context of
various scenarios for their existence: as tidal features, as galaxies
with extremely low surface brightness stellar populations, or as
genuine ``dark'' galaxies.  The ALFALFA ``Almost Dark'' objects fall
into four broad categories: 1) objects with nearby \HI\ neighbors that
are likely of tidal origin; 2) objects that appear to be part of a
system of multiple \HI\ sources, but which may not be tidal in origin;
3) objects isolated from nearby ALFALFA \HI\ detections, but located
near a gas-poor early-type galaxy; 4) apparently isolated sources,
with no object of coincident redshift within $\sim$400 kpc.  The five
sources presented here (AGC\,193953, AGC\,208602, AGC\,208399,
AGC\,226178, and AGC\,233638) populate each of the last three
categories (sources with a clear tidal origin from ALFALFA data
products were not targeted for interferometric follow-up).
Table~\ref{table_basic} summarizes the basic properties of the five
sources, including position, distance, and \HI\ properties
(derived from ALFALFA measurements).  This limited sample is not meant
to be representative of the ``Almost Dark'' galaxy sample as a whole;
these specific sources were selected in part to fit into pre-approved
LST observing slots for VLA scheduling (see \S~\ref{S2.1}).  

There has been widespread interest in identifying truly ``dark''
galaxies, if they exist.  With the advent of sensitive, wide-area
\HI\ surveys such as ALFALFA, and with the increased availability of
mapping observations of nearby groups with single dish telescopes, a
number of candidate ``dark galaxies'' have appeared in the literature.
These are systems whose \HI\ properties are representative of
galaxies, but that lack a clear optical counterpart.  The
VIRGO\,HI\,21 cloud was originally postulated to be a true dark galaxy
\citep{minchin05,minchin07}.  However, subsequent observations
\citep{haynes07} and modeling \citep{duc08} have shown that this source
is likely tidal in nature.  More recently, single-dish telescopes have
been used in mapping mode to produce deep targeted \HI\ images of
nearby groups.  In such a survey of the M\,101 group, \citet{mihos12}
discovered a gas cloud labeled as GBT\,1355$+$5439.  In that same
work, very deep ground-based optical imaging revealed no obvious
stellar counterpart to a limiting surface brightness of $\mu_{\rm V}$
$=$ 29 mag\,arcsec$^{-2}$.  Subsequent interferometric
\HI\ observations by \citet{oosterloo13} confirm the extreme
characteristics of GBT\,1355$+$5439; both a tidal interpretation and
an interpretation as a truly ``dark'' object remain viable.  Numerous
other sources are interesting and relevant in this context (see
discussion and references in {Walter \etal\ 2005}\nocite{walter05},
{Kent \etal\ 2009}\nocite{kent09}, {English
  \etal\ 2010}\nocite{english10}, {Kent 2010}\nocite{kent10},
{Matsuoka \etal\ 2012}\nocite{matsuoka12}, {Serra
  \etal\ 2013}\nocite{serra13}, {Lee-Waddell
  \etal\ 2014}\nocite{leewaddell14}, {Adams
  \etal\ 2014}\nocite{adams14}).

We organize this manuscript as follows.  In \S~\ref{S2} we discuss the
details of the \HI\ data reductions and image production, as well as
the techniques applied to derive optical and ultraviolet (UV)
properties from SDSS and GALEX images.  \S~\ref{S3} presents the \HI,
optical, and UV images of each of the five sources; using these data,
we can localize the \HI\ gas with confidence, assign possible optical
counterparts (by comparison with optical imaging), and search for
ordered motions that are indicative of galaxy-like rotation.  In
\S~\ref{S4} we further discuss the five ``Almost Dark'' sources in the
context of the ALFALFA database, the low-mass galaxies within it, and
also among the (growing) list of ``Almost Dark'' candidates.  We draw
our conclusions in \S~\ref{S5}; primary among these is that none of
the five candidate objects appears to qualify as a bona fide ``dark
galaxy''.

\section{Observations and Data Handling}
\label{S2}
\subsection{HI Observations}
\label{S2.1}

\HI\ spectral line imaging of the five sources presented in this work
was obtained with the National Radio Astronomy Observatory's {\it Karl
  G. Jansky Very Large Array} ({\it VLA}\footnote{The National Radio
  Astronomy Observatory is a facility of the National Science
  Foundation operated under cooperative agreement by Associated
  Universities, Inc.}) in April, 2013, for program 13A-028 (legacy ID
AC1116; P.I. Cannon).  The five sources selected for observing were
the best fits based on the LST oversubscription of the VLA during the
13A semester.  The observations were acquired during four
separate two-hour observation blocks and during one 2.5-hour block
(for source AGC\,193953) with the observatory in the ``D'' (most
compact) configuration.  The WIDAR correlator was used to divide an 8
MHz bandwidth into 1024 channels.  The resulting channel separation of
7.812 kHz\,ch$^{-1}$ delivers a native spectral resolution of 1.65
km\,s$^{-1}$\,ch$^{-1}$ at the rest frequency of the \HI\ spectral
line.  Reductions followed standard prescriptions and used both the
Astronomical Image Processing System (AIPS) and the Common Astronomy
Software Applications (CASA) packages.

Radio frequency interference (RFI) and bad data were removed manually
from each dataset.  The RFI contamination of the AGC\,193953 dataset
was significant in a few channels close to those containing the
source; the loss of sensitivity in these channels has affected our
recovery of flux from this source (see further discussion below).
Bandpass and flux calibrations used the standard calibrator 3C286 for
all five observing blocks.  For each source, a nearby quasar was
observed every $\sim$20 minutes and used to calibrate the phases. We
estimate our absolute flux calibration to be accurate at the
$\sim$10\% level.

The imaging of the calibrated $uv$ databases followed standard
prescriptions.  To increase S/N, each database was spectrally averaged
by a factor of two, to produce a velocity resolution of 15.625
kHz\,ch$^{-1}$ (3.3 km\,s$^{-1}$\,ch$^{-1}$) prior to imaging.  Each
cube was inverted with minimal cleaning, and source(s) were
identified.  Using these full cubes, regions were identified that
excluded \HI\ line sources and bandpass edge effects.  Using
these regions (typically 50 channels or more, depending on the
complexity of emission within each cube) as fitting points, the
continuum was then subtracted in the $uv$ plane using a first-order
polynomial function.

The final spectral line cubes used in this analysis were produced
using the IMAGR task in AIPS, with a weighting ROBUST factor of 0.5.
Using clean boxes that tightly encompass the high S/N line emission,
each cube was deeply cleaned to one half of the RMS level found in
line-free channels.  Residual flux rescaling was enforced during the
imaging process \citep{jorsater95}.  The moment zero maps shown below
were produced following the THINGS procedures as described in
\citet{walter08}.  The original-resolution cubes were convolved to the
nearest integer circular sizes in arcseconds (so, for example, the
48.97\arcsec$\times$43.00\arcsec\ cube for AGC\,208602 was convolved
to a circular beam size of 49\arcsec) and were then used for all
subsequent analyses.  A second convolution was enacted to produce
cubes with circular beam sizes $\sim$10\% larger; these smoothed cubes
were blanked at the 2.5$\sigma$ level, closely examined by eye to
retain only emission that is coherent in velocity space, and then used
as masks against the original (circular beam size) cubes.  Moment zero
maps (representing total \HI\ emission) were produced by collapsing
these masked cubes.  The resulting images of each source are discussed
in detail below.  The total \HI\ flux integral for each of the five
sources is less than the flux integral derived from the ALFALFA
observations; we discuss the recovered flux from each source in the
sections that follow.

The velocity structure of the \HI\ gas in each source plays an
important role in the interpretation of these data; the presence of
ordered gas motion is very strong evidence for classifying a candidate
``Almost Dark'' source as a bona-fide galaxy.  As discussed in detail
for each individual system below, whenever possible we have used
Gaussian fitting to the full (unblanked) datacubes in order to extract
the velocity fields of the ``Almost Dark'' systems presented below.
Specifically, we use the GIPSY task XGAUFIT to fit Gaussian functions
to emission above the 3$\sigma$ level in regions of each datacube
immediately surrounding 4 of the 5 ``Almost Dark'' sources; these
Gaussian-fitted velocity fields are shown in the 4-panel images
presented in \S~\ref{S3}.  For AGC\,208602, the incoherent velocity
structure of the source precluded meaningful Gaussian fitting.  For
those ``Almost Dark'' sources that have neighboring \HI-rich galaxies
that are detected in the datacubes (AGC\,208602, AGC\,226178, and
AGC\,233638), we also show full-field images including all detected
galaxies using the more traditional moment one approach (i.e.,
collapsing of the blanked data cube) for creating the velocity fields.

\subsection{Optical Observations}
\label{S2.2}

To quantify the optical properties of these five ``Almost Dark''
sources, we extract photometry from the Sloan Digital Sky Survey
10$^{\rm th}$ data release products \citep{ahn14}.  Mosaic frames in
the SDSS-g and the SDSS-r bands, using primary covering fields, were
created using the SDSS online interface.  Integrated magnitudes were
derived for each of the ``Almost Dark'' sources using the LAMBDAR
source extraction code (Wright \etal, in preparation\footnote{For
  details see https://github.com/AngusWright/LAMBDAR}).  Elliptical
apertures, centered on the \HI\ centroid positions (see
Table~\ref{table_basic}) and with semi-major and semi-minor axis
lengths as shown in Table~\ref{table_observed}, were used to extract
the SDSS-g magnitudes and (g$-$r) colors of each source.  The aperture
sizes and orientations were set by hand in order to enclose all of the
optical flux from each source when the SDSS images were displayed at
the highest contrast levels.

The photometry in Table~\ref{table_observed} is used to derive the
physical parameters of the five ``Almost Dark'' sources shown in
Table~\ref{table_derived}.  After correcting for foreground Galactic
absorption \citep{schlafly11}, the SDSS g magnitude is converted to a
Johnson-Cousins B magnitude via the prescriptions in Table 2 of
\citet{blanton07}.  These (absorption corrected) apparent magnitudes
are converted to absolute magnitudes using our assumed distances from
Table~\ref{table_basic}, and then converted to solar luminosities
using the values from Table 1 of \cite{blanton07}.  The resulting g
and B-band luminosities of each ``Almost Dark'' source are tabulated
and compared with the total \HI\ masses (using the ALFALFA flux
integrals) in Table~\ref{table_derived}.  This table reveals that the
M$_{\rm HI}$/L ratios span more than a factor of 10 across this small
sample.  It is important to stress that all of our measurements of the
optical properties of these sources are brighter than the pipeline
SDSS values; this is a result of the more careful aperture placement
and background subtraction done here than by the standard SDSS
pipeline measuring tools.

We use the measured r-band magnitude, (g$-$r) color, and the relations
from \citet{zibetti09} to derive the stellar mass and M$_{\rm
  HI}$/M$_{\star}$ ratio of each candidate ``Almost Dark'' source
shown in Table~\ref{table_derived}.  We stress that these
representative stellar masses are uncertain at the $\sim$50\% level
due to systematic uncertainties alone.  Nonetheless, these values can
be compared to those for the full $\alpha$.40 catalog using the same
technique; we explore this comparison later in this manuscript.

A detailed investigation of the optical properties of the ALFALFA
``Almost Darks'' sample is currently underway with the One Degree
Imager (ODI\footnote{http://www.noao.edu/wiyn/ODI/}) on the WIYN 3.5m
telescope\footnote{http://www.noao.edu/wiyn/}.  The results of this
endeavor will be presented in a forthcoming manuscript (Janowiecki
\etal, in preparation).  As of this writing, preliminary images of two
members of the present sample (AGC\,193953 and AGC\,233638) have been
obtained as part of this effort.  In the interests of uniformity, we
only derive quantitative results for the members of the present pilot
sample from the SDSS images; the preliminary derivations of SDSS
g-band magnitudes for AGC\,193953 and AGC\,233638 from the partially
filled focal plane ODI (pODI) images agree within errors with the
values derived from the original SDSS images.

\subsection{Ultraviolet Observations}
\label{S2.3}

The ultraviolet (UV) properties of the five ``Almost Dark'' sources
were derived using archival GALEX\footnote{Based on observations made
  with the NASA Galaxy Evolution Explorer.  GALEX is operated for NASA
  by the California Institute of Technology under NASA contract
  NAS5-98034.} imaging.  The depth of the GALEX images is
heterogeneous across the sample; two sources (AGC\,193953 and
AGC\,233638) have only All-Sky Imaging Survey depth imaging ($\sim$100
sec per integration), and one of these (AGC\,193953) is a
non-detection.  The other three sources have deeper imaging
($\sim$1600 sec per integration); one source (AGC\,208602) is a
non-detection at this sensitivity level.  Table~\ref{table_observed}
shows the magnitudes in the GALEX near and far ultraviolet channels,
measured in identical apertures to those used to extract the optical
photometry shown in Table~\ref{table_observed}.  We use the far-UV
fluxes to derive the recent ($\sim$100 Myr) star formation rates
(SFR$_{\rm FUV}$) of the sources using the prescriptions in
\citet{salim07}; these quantities are discussed in detail below.

\section{The Gaseous and Stellar Components of the ``Almost Dark'' Sources}
\label{S3}

In this section we present the \HI, optical, and UV images of each of
the ``Almost Dark'' sources.  We group these five systems into three
categories: isolated systems (no known sources with coincident
redshift within a projected distance of 400 kpc), systems with an
apparent tidal origin, or systems near other objects.

\subsection{Isolated Sources}
\label{S3.1}

We classify two of the five ``Almost Dark'' sources as isolated; no
known sources with coincident redshift are located within a projected
distance of 400 kpc of either source.

\subsubsection{AGC\,193953}
\label{S3.1.1}

AGC\,193953 has the weakest ALFALFA \HI\ line flux in this small
sample (S$_{\rm HI}$ $=$ 0.43 Jy\,km\,s$^{-1}$; see
Table~\ref{table_basic}).  Located at $\sim$40 Mpc, the \HI\ line
width (W$_{\rm 50}$ $=$ 32 \kms, where W$_{\rm 50}$ is the width of
the \HI\ line at 50\% of maximum) and \HI\ reservoir of
1.6\,$\times$\,10$^{8}$ \msun\ are indicative of an average dwarf
galaxy.  This source was initially cataloged as a ``dark'' candidate
due to the extremely low surface brightness nature of the optical
counterpart, which is significantly displaced from the center of the 
ALFALFA beam.  As discussed above, RFI compromised a number of channels
in the datacube close to those that contain emission from the source.
As such, the total \HI\ flux in the VLA data (0.13 Jy\,km\,s$^{-1}$)
only amounts to $\sim$30\% of the ALFALFA flux; the \HI\ properties
(e.g., the more narrow range of velocities detected in the VLA data
compared to the ALFALFA data) should only be considered to be
representative based on these data.

\begin{figure}
\epsscale{1}
\plotone{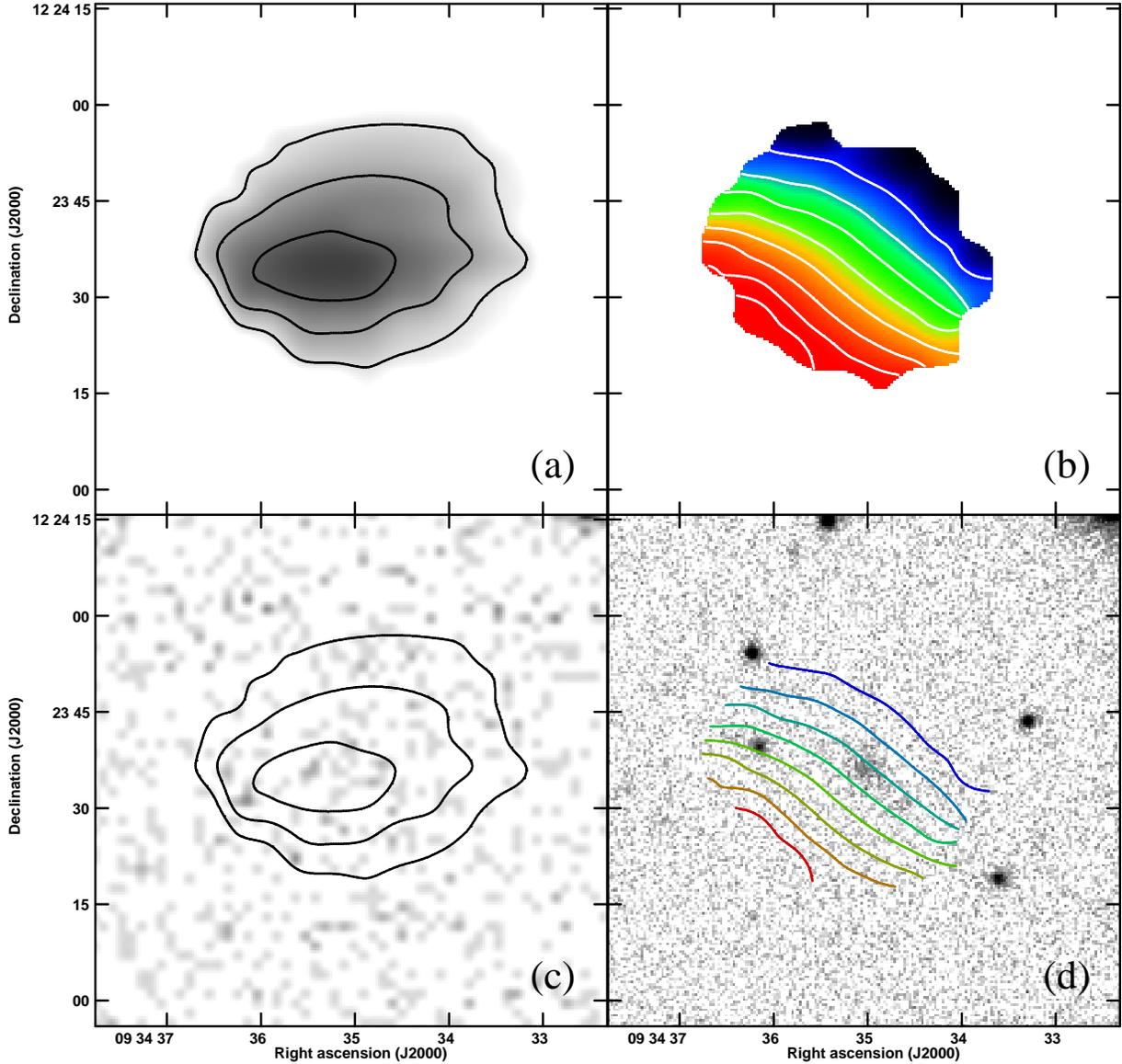}
\caption{\HI, SDSS R-band, and GALEX images of AGC\,193953.  Panel (a)
  shows \HI\ column density contours at the (0.4, 0.5,
  0.6)\,$\times$\,10$^{20}$ cm$^{-2}$ levels, overlaid on a greyscale
  representation of the image.  Panel (b) shows \HI\ isovelocity
  contours from 2596 km\,s$^{-1}$ to 2603 km\,s$^{-1}$ in intervals of
  1 km\,s$^{-1}$, overlaid on a color representation of the velocity
  field image.  Panel (c) shows the GALEX near-UV image, while panel
  (d) shows the SDSS r-band image (displayed in a logarithmic stretch
  to highlight low surface brightness emission); the contours in (c)
  are the same as in panel (a), while the contours in (d) are the same
  as in panel (b).  The velocity field was derived by fitting Gaussian
  functions to the full data cube.  The \HI\ beam size is 50\arcsec;
  the source is unresolved at this angular resolution.}
\label{images-193953}
\end{figure}

Figure~\ref{images-193953} shows the \HI, SDSS-r band, and GALEX
near-UV images of AGC\,193953.  At D configuration angular resolution
the source is unresolved.  The \HI\ gas has a single surface density
concentration that reaches a maximum column density of only
6.5\,$\times$\,10$^{19}$ cm$^{-2}$.  This surface density maximum is
slightly offset spatially (though within one beam size) from the very
low surface brightness optical counterpart, which is clearly visible
in the SDSS-r band image.  The Gaussian-fitted velocity field shows
coherent and ordered rotation within the beam.  As panels (b) and (d)
of Figure~\ref{images-193953} show, the isovelocity contours are
parallel over $\sim$7 \kms\ ($\sim$2 velocity resolution elements).

The optical counterpart of AGC\,193953 is of very low surface
brightness.  As Tables~\ref{table_observed} and \ref{table_derived}
show, the SDSS integrated magnitudes imply a total B-band luminosity
of $\sim$1.9\,$\times$\,10$^{7}$ L$_{\odot}$ and a total stellar mass
of 2.2\,$\times$\,10$^{6}$ \msun.  This faint stellar component is
detected in the shallow GALEX near-UV image; while we have no direct
constraints on the nature of recent star formation (which requires
far-UV imaging), the faint nature of the source in the near-UV argues
for quiescence over the most recent $\sim$100 Myr.  Using the ALFALFA
total \HI\ flux and the SDSS-g and derived B-band luminosities, we
derive M$_{\rm HI}$/L$_{\rm g}$ $=$ 10, M$_{\rm HI}$/L$_{\rm B}$ $=$
9, and M$_{\rm HI}$/M$_{\star}$ $=$ 75.  While our analysis of the
SDSS images have recovered more flux than the SDSS pipeline products,
the source is close to the detection limits of the SDSS imaging; it is
likely that deeper optical imaging would recover more low surface
brightness emission, thus decreasing the M$_{\rm HI}$/L and M$_{\rm
  HI}$/M$_{\star}$ ratios.

\subsubsection{AGC\,208399}
\label{S3.1.2}

AGC\,208399 is a very low surface brightness source that was included
in the {Huang \etal\ (2012)}\nocite{huang12} study of gas, stars, and
star formation in ALFALFA dwarfs.  AGC\,208399 has the lowest
recessional velocity of the five ``Almost Dark'' sources studied here;
its flow model distance (see Table~\ref{table_basic}) is 10.1 Mpc.
The flux integral (S$_{\rm HI}$ $=$ 1.06 Jy\,km\,s$^{-1}$) implies an
\HI\ mass of 2.6\,$\times$\,10$^{7}$ \msun.  By comparison with the
properties in the {Huang \etal\ (2012)}\nocite{huang12} sample,
AGC\,208399 is an extreme low surface brightness dwarf galaxy.

\begin{figure}
\epsscale{1}
\plotone{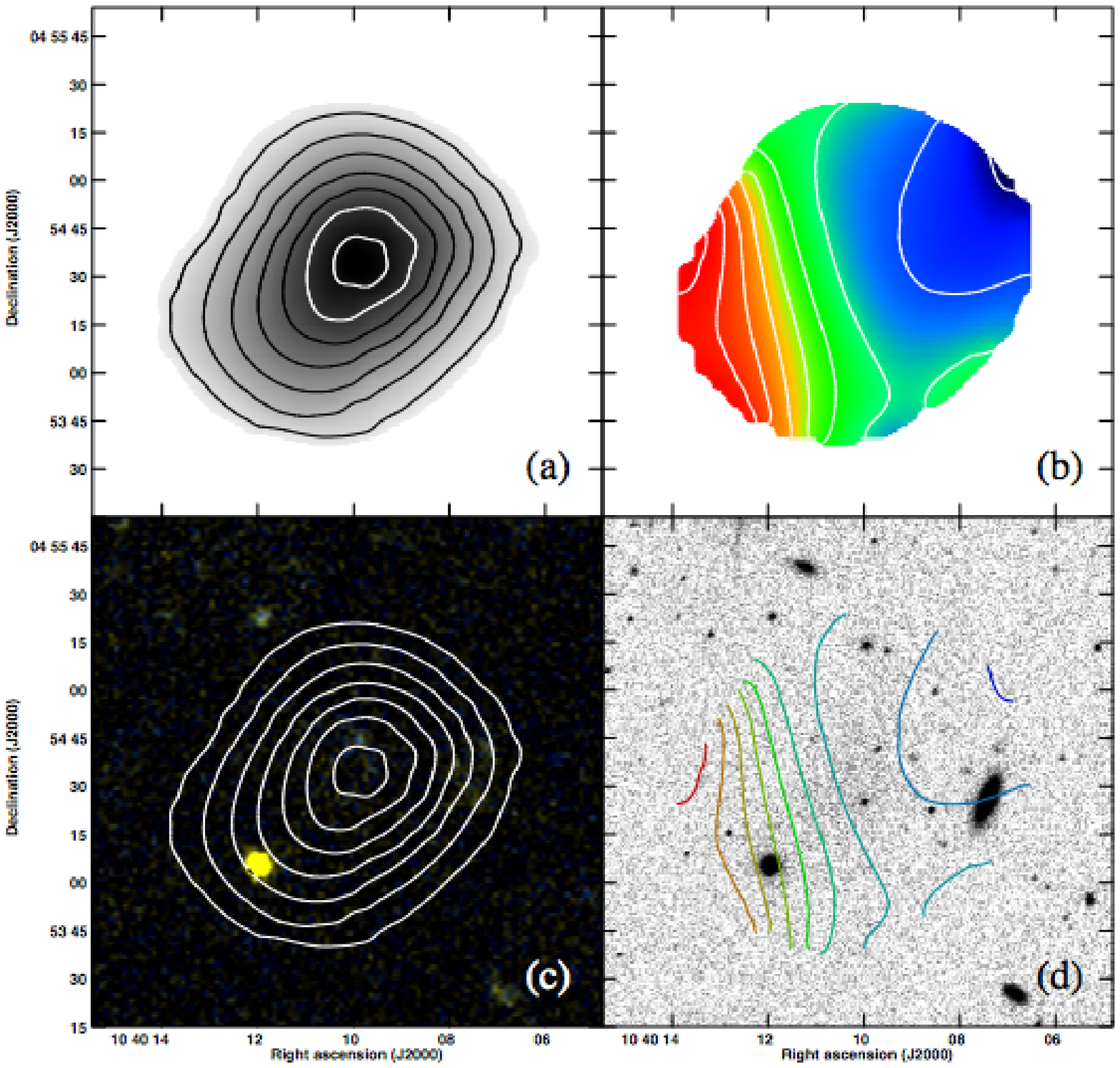}
\caption{\HI, SDSS R-band, and GALEX images of AGC\,208399.  Panel (a)
  shows \HI\ column density contours at the (0.5, 0.75, 1.0, 1.25,
  1.5, 1.75, 2.0)\,$\times$\,10$^{20}$ cm$^{-2}$ levels, overlaid on a
  greyscale representation of the image.  Panel (b) shows
  \HI\ isovelocity contours from 760 km\,s$^{-1}$ to 776 km\,s$^{-1}$
  in intervals of 2 km\,s$^{-1}$, overlaid on a color representation
  of the velocity field image.  Panel (c) shows a color representation
  of the GALEX UV images, while panel (d) shows the SDSS r-band image
  (displayed in a logarithmic stretch to highlight low surface
  brightness emission); the contours in (c) are the same as in panel
  (a), while the contours in (d) are the same as in panel (b).  The
  velocity field was derived by fitting Gaussian functions to the full
  data cube.  The \HI\ beam size is 52\arcsec; the source is
  marginally resolved at this angular resolution.  The two galaxies in
  panel (d) located close to AGC\,208399 (to the W and SW) are
  background galaxies with SDSS redshifts of 0.065 and 0.071,
  respectively.}
\label{images-208399}
\end{figure}

The \HI, optical and UV images of AGC\,208399 shown in
Figure~\ref{images-208399} reveal a centrally concentrated
\HI\ distribution that is spatially coincident with a diffuse, very
low surface brightness optical counterpart in the SDSS r-band image.
The VLA images recover $\sim$85\% of the ALFALFA flux integral.  The
\HI\ column density peaks at 2.1\,$\times$\,10$^{20}$ cm$^{-2}$ at
52\arcsec\ angular resolution, which marginally resolves the source.
There is bulk rotation of the \HI\ gas, extending across nearly 20
\kms.  Most of this projected rotation is coherent, as evidenced by
the nearly parallel isovelocity contours over most of the disk.
However, there are some departures from circular motion in the outer
regions, especially on the approaching side of the disk.

The GALEX imaging shown in Figure~\ref{images-208399}(c) shows a very
low surface brightness blue component in close angular proximity to
the \HI\ surface density maximum.  It is difficult to discern if this
faint emission arises from the stellar population associated with
AGC\,208399 (and thus likely indicates recent star formation within
the past $\sim$100 Myr) or with unrelated background sources; indeed,
the SDSS-r image shown in panel (d) of Figure~\ref{images-208399}
reveals some small point sources that could be background star-forming
galaxies.  With this caution in mind, the UV fluxes are extracted
using the same apertures as for the SDSS photometry (see
Table~\ref{table_observed}).  If the far-UV flux is associated with
AGC\,208399, then the implied SFR$_{\rm FUV}$ $=$
5\,$\times$\,10$^{-5}$ \msun\,yr$^{-1}$.

AGC\,208399 shares many physical characteristics with the most massive
dwarf galaxies in SHIELD \citep{cannon11}, a complementary ALFALFA
follow-up observing campaign to study the physical properties of
extremely low-mass systems cataloged by ALFALFA.  Specifically, the
SDSS colors, \HI\ line widths, total \HI\ masses, and distances of the
most massive SHIELD galaxies are similar to those of AGC\,208399.
Interestingly, however, the majority of the SHIELD galaxies show
evidence for recent star formation.  The gas-rich (M$_{\rm
  HI}$/L$_{\rm g}$ $=$ 6; M$_{\rm HI}$/L$_{\rm B}$ $=$ 5; M$_{\rm
  HI}$/M$_{\star}$ $=$ 15) but quiescent (SFR$_{\rm FUV}$ $=$
5\,$\times$\,10$^{-5}$ \msun\,yr$^{-1}$) nature of AGC\,208399
warrants further detailed exploration.

\subsection{Systems With A Tidal Origin}
\label{S3.2}

\subsubsection{AGC\,208602}
\label{S3.2.1}

AGC\,208602 was classified as a source with \HI\ in the vicinity of,
but offset from, an early-type galaxy (NGC\,3457).  A comparison of
the ALFALFA detections near AGC\,208602 with SDSS imaging revealed
that this candidate ``Almost Dark'' source is located within
30\arcmin\ of at least four known \HI\ sources, including NGC\,3443,
NGC\,3454, NGC\,3455, and UGC\,6022.  This loose collection is part of
the NGC\,3370 group of galaxies, which is in turn part of the larger
Leo\,II structure, which contains multiple gas-rich spirals and
irregular galaxies.  It also contains at least one early-type galaxy;
NGC\,3457 is in close angular proximity to AGC\,208602, and its
optical velocity (V$_{\rm sys}$ = 1148 \kms; {Cappellari
  \etal\ 2011}\nocite{cappellari11}) is in fair agreement with those
of other members of the group (see below).

Figure~\ref{images-208602-full} shows VLA \HI\ and SDSS optical images
of the region surrounding the ``Almost Dark'' candidate, AGC\,208602.
The size of the field of view ($\sim$36\arcmin) is roughly equal to
that of the primary beam of the VLA at \HI.  Within this field, five
other \HI\ line sources are detected in the velocity range immediately
surrounding AGC\,208602 (specifically, between $\sim$985 \kms\ and
$\sim$1210 \kms); these five sources are noted in each panel of
Figure~\ref{images-208602-full}.  In addition to NGC\,3443, NGC\,3454,
NGC\,3455, and UGC\,6022, we also detect the neutral hydrogen in a
small system labeled as SDSS\,J105506.66$+$172745.4.  Each of these
five sources shows coherent signatures of rotation.

\begin{figure}
\epsscale{1}
\plotone{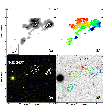}
\caption{\HI, SDSS R-band, and GALEX images of AGC\,208602.  Panel (a)
  shows \HI\ column density contours at the (0.35, 0.45,
  0.55)\,$\times$\,10$^{20}$ cm$^{-2}$ levels, overlaid on a greyscale
  representation of the image.  Panel (b) shows \HI\ isovelocity
  contours from 1085 km\,s$^{-1}$ to 1130 km\,s$^{-1}$ in intervals of
  5 km\,s$^{-1}$, overlaid on a color representation of the velocity
  field image.  Panel (c) shows a color representation of the GALEX UV
  images, while panel (d) shows the SDSS r-band image (displayed in a
  logarithmic stretch to highlight low surface brightness emission);
  the contours in (c) are the same as in panel (a), while the contours
  in (d) are the same as in panel (b). The early-type galaxy NGC 3457
  is labeled in panel (c); the systemic velocity of this galaxy (V $=$
  1148 \kms; Cappellari \etal\ 2011) suggests a physical association
  with the other galaxies in the AGC\,208602 field.  The velocity
  field was derived via a standard first moment analysis, using the
  same pixels that contribute to the column density (moment zero)
  image; the incoherent velocity structure of AGC\,208602 precluded
  meaningful Gaussian fitting to the full data cube.  The \HI\ beam
  size is 49\arcsec, as shown in panel (a).}
\label{images-208602}
\end{figure}

Figure~\ref{images-208602} shows a closer view of the detected
\HI\ from AGC\,208602, and its relationship to NGC\,3457.  The neutral
hydrogen in this source is of very low surface brightness,
characterized by flocculent patches of emission that are at times
disjoint in velocity space.  While the VLA images recover $\sim$65\%
of the ALFALFA flux, the integrated column densities at
49\arcsec\ resolution only reach the $\sim$5.5\,$\times$\,10$^{19}$
cm$^{-2}$ level.  The velocity structure of this source is
sufficiently incoherent that attempts to fit Gaussians to the emission
in the cube failed.  Thus, the moment one maps shown in
Figures~\ref{images-208602-full} and ~\ref{images-208602} were derived
by standard routines.  Emission is detected over $\sim$50 \kms; there
is no convincing evidence of coherent rotation of this source in the
velocity field or in the 3-dimensional cube.

The SDSS and GALEX images of AGC\,208602 presented in
Figure~\ref{images-208602} reveal no detectable optical counterpart.
The magnitudes derived in Table~\ref{table_observed} are thus listed as
limits; the corresponding luminosities, M$_{\rm HI}$/L ratios
(M$_{\rm HI}$/L$_{\rm g}$ $>$ 48; M$_{\rm HI}$/L$_{\rm B}$ $>$ 45),
and M$_{\rm HI}$/M$_{\star}$ ratio ($>$49) should likewise be
considered limiting values.  Deeper optical imaging of this source might
be fruitful, but the interpretation may be complicated by the outer
halo of NGC\,3457.

AGC\,208602 stands out markedly in Table~\ref{table_derived}; it has
the largest M$_{\rm HI}$/L ratios of any of the five sources in the
present work.  Further, the irregular morphology and incoherent
velocity structure are unlike those found in any of the other systems
studied here or in other low-mass galaxies.  Based on these factors,
and on the source's membership in the NGC\,3370 Group, we conclude
that this feature may have a tidal origin.  It shares some physical
characteristics with other ALFALFA-detected groups with extended tidal
structures (e.g., {Lee-Waddell \etal\ 2014}\nocite{leewaddell14} and
references therein).

It is not clear which sources in the NGC\,3370 Group might have
interacted to give rise to this object.  At the adopted distance of
18.5 Mpc, the 1.2\,$\times$\,10$^{8}$ \msun\ of \HI\ associated with
AGC\,208602 spans more than 20 kpc.  This remarkable cloud is located
roughly 65 kpc from UGC\,6022 and SDSS\,J105506.66$+$172745.4, the two
gas-rich systems with the closest angular projection on the sky.
Further, AGC\,208602 is located between NGC\,3443 ($\sim$126 kpc
projected separation) and the early-type galaxy NGC\,3457 ($\sim$15 kpc
projected separation); it may be the case that the feature was
associated with NGC\,3443, and that a recent close encounter with
NGC\,3457 has stripped the \HI\ away from NGC\,3443.  A very deep
single-dish map of the entire NGC\,3370 Group might be fruitful; this
would provide sensitivity to possible low surface brightness tidal
structure.

\subsection{Systems Near Other Galaxies}
\label{S3.3}

The remaining two ``Almost Dark'' sources are located in sufficiently
close angular proximity to other known sources that the assignment of
an unambiguous optical counterpart was not possible using the ALFALFA
data alone.

\subsubsection{AGC\,226178}
\label{S3.3.1}

AGC\,226178 is an \HI\ source with no optical counterpart at the
ALFALFA centroid.  The system is a possible member of the Virgo
Cluster; \citet{kent08} catalog its \HI\ properties, but find no
optical counterpart.  They note the possible association with
AGC\,229166, an optically faint, low surface brightness object (not
detected in \HI\ by ALFALFA, but cataloged as an AGC galaxy based on
the extended optical emission which was apparent when examining the
ALFALFA data) located at ($\alpha$,$\delta$) $=$
(12$^{h}$46$^{m}$41.7$^{s}$, 10\arcdeg23\arcmin09\arcsec), roughly
2.3\arcmin\ from the ALFALFA centroid for AGC\,226178.  The VLA images
presented here show the \HI\ centroid to be 1\arcmin\ south of
AGC\,229166, and suggest that the \HI\ is associated with a second
possible optical/UV counterpart with an irregular morphology.

\begin{figure}
\epsscale{0.75}
\plotone{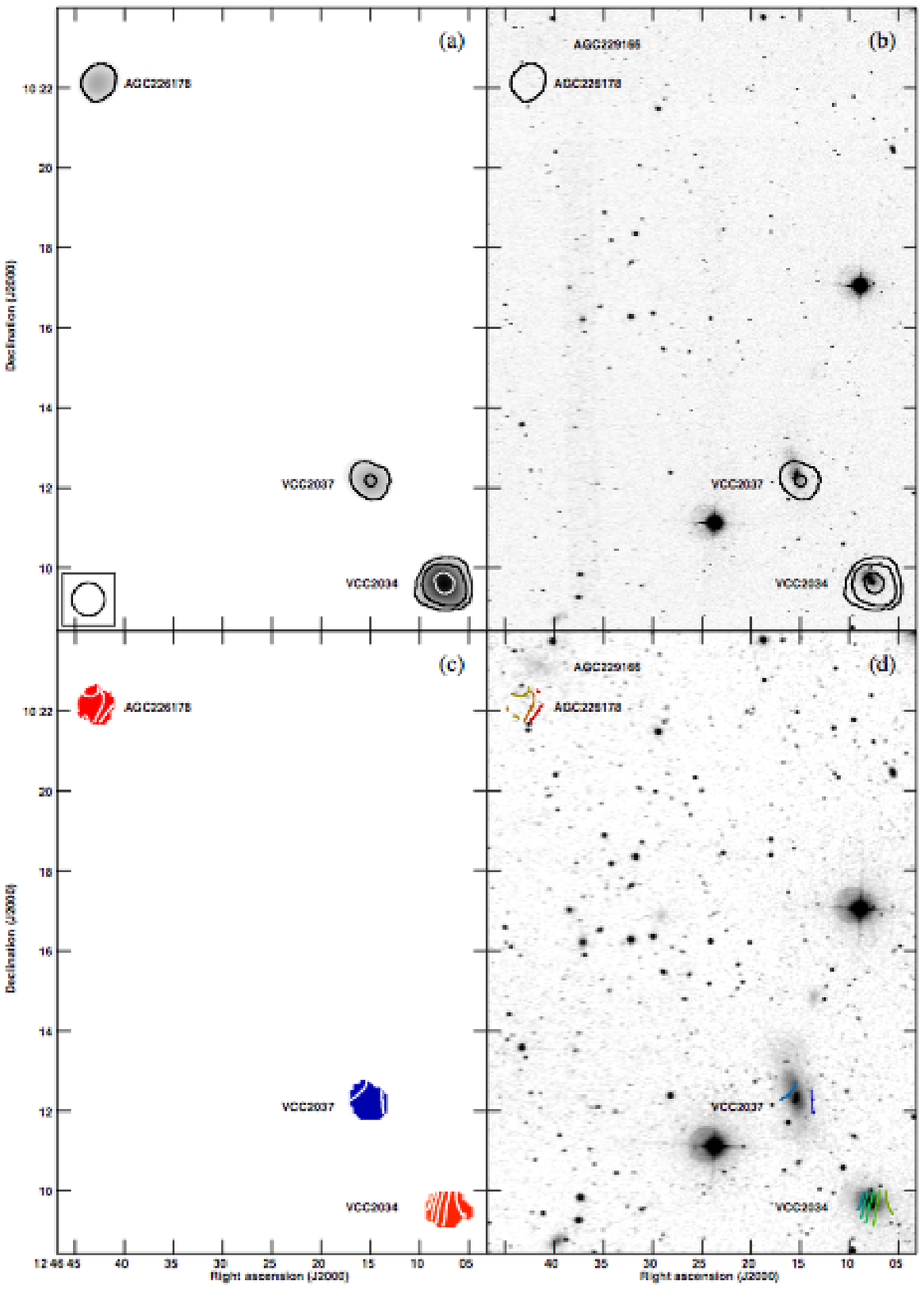}
\caption{\HI\ and SDSS R-band images of the AGC\,226178 field.  Panel
  (a) shows \HI\ column density contours at the (0.5, 1,
  2)\,$\times$\,10$^{20}$ cm$^{-2}$ levels, overlaid on a greyscale
  representation of the image.  Panel (c) shows \HI\ isovelocity
  contours at the (1145, 1148, 1495, 1500, 1505, 1510, 1515, 1586,
  1588, 1590) \kms\ levels overlaid on a color representation of the
  image; the color bar is set to encompass the full velocity range of
  \HI\ gas in the three detected galaxies.  Panels (b) and (d) show
  the SDSS r-band image of the AGC\,226178 field; panel (b) is
  displayed in a logarithmic stretch to highlight low surface
  brightness emission, while panel (d) has been smoothed by a Gaussian
  kernel (see discussion in \S~\ref{S3.2}).  The contours in (b) are
  the same as in panel (a), while the contours in (d) are the same as
  in panel (c).  The very low optical surface brightness source
  AGC\,229166 is also labeled in panels (b) and (d).  The \HI\ beam
  size is 49\arcsec, as shown in the bottom left of panel (a).}
\label{images-226178-full}
\end{figure}

AGC\,226178 is in close angular and velocity proximity to other
\HI-bearing sources in our spectral line data.  As shown in
Figure~\ref{images-226178-full}, two other Virgo dwarf galaxies are
detected at high significance: VCC\,2034 (AGC\,221001; V$_{\rm sys}$
$=$ 1504 \kms; angular separation $\simeq$910\arcsec\ $\simeq$72.8
kpc) and VCC\,2037 (AGC\,221004; V$_{\rm sys}$ $=$ 1142 \kms; angular
separation $\simeq$720\arcsec\ $\simeq$57.5 kpc).  We do not find
clear evidence for an ongoing interaction between any of these sources
at our current sensitivity level.

\begin{figure}
\epsscale{1}
\plotone{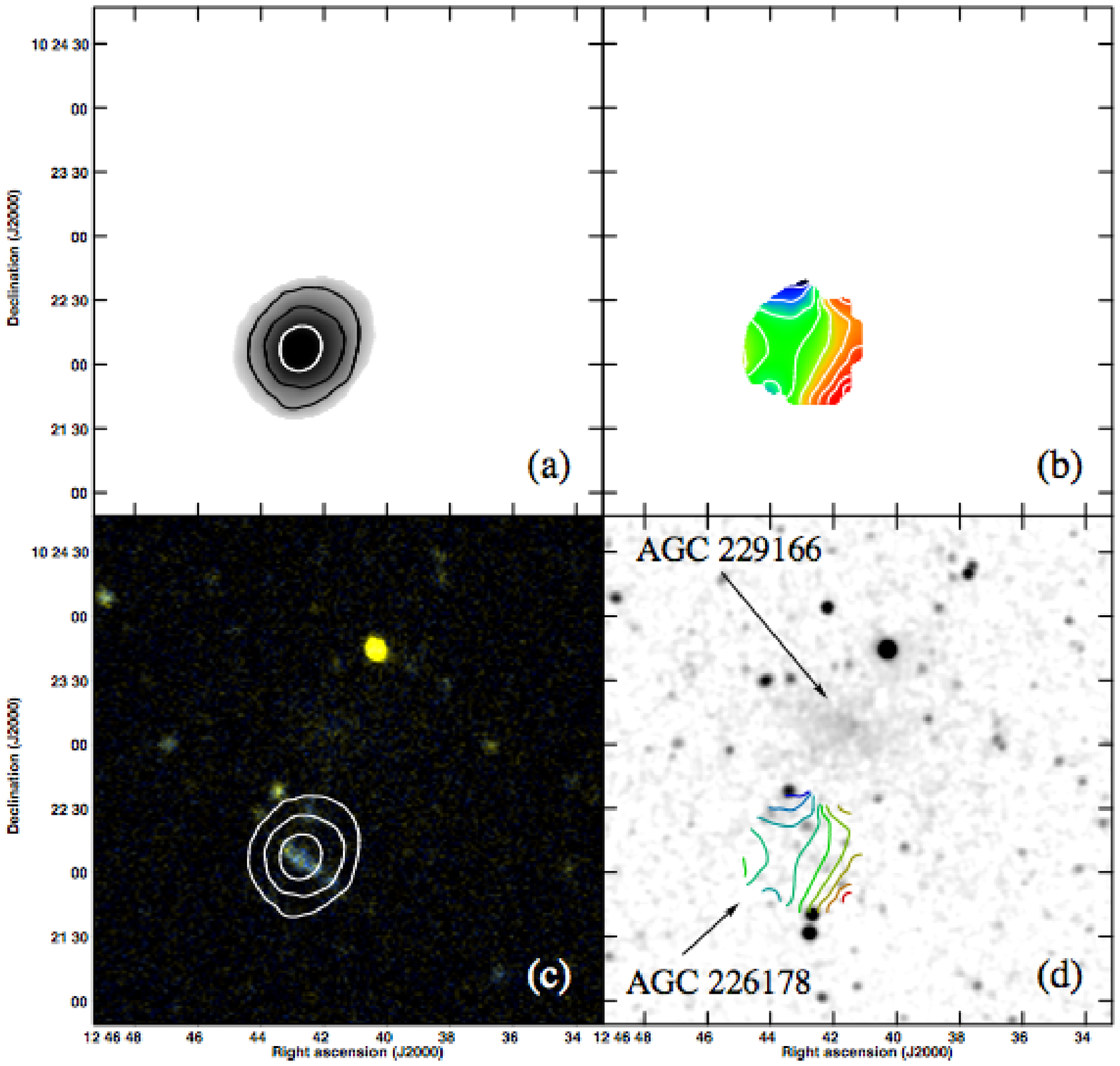}
\caption{\HI\ and SDSS R-band images of AGC\,226178.  Panel (a) shows
  \HI\ column density contours at the (0.5, 0.7,
  0.9)\,$\times$\,10$^{20}$ cm$^{-2}$ levels, overlaid on a greyscale
  representation of the image.  Panel (b) shows \HI\ isovelocity
  contours from 1584 km\,s$^{-1}$ to 1592 km\,s$^{-1}$ in intervals of
  1 km\,s$^{-1}$, overlaid on a color representation of the velocity
  field image.  Panel (c) shows a color representation of the GALEX UV
  images, while panel (d) shows the SDSS r-band image (convolved by a
  Gaussian kernel to highlight low surface brightness emission); the
  contours in (c) are the same as in panel (a), while the contours in
  (d) are the same as in panel (b).  The velocity field was derived by
  fitting Gaussian functions to the full data cube.  The \HI\ beam
  size is 49\arcsec; the source is unresolved at this angular
  resolution.}
\label{images-226178}
\end{figure}

At SDSS imaging depths, the optical counterpart of the AGC\,226178
\HI\ source is ambiguous.  To highlight the low surface brightness
nature of both AGC\,226178 and AGC\,229166, panel (d) of
Figure~\ref{images-226178-full} shows the SDSS-r band image from panel
(b), smoothed by a Gaussian kernel with $\sigma$ $=$ 3 pixels.  The
\HI\ is coincident with a very faint irregular optical counterpart
candidate, but offset from AGC\,229166.  Figure~\ref{images-226178}
shows the \HI, optical, and UV properties in a smaller region
containing only these two sources; note that the SDSS-r band image in
this figure is the same as shown in panel (d) of
Figure~\ref{images-226178-full} (i.e., it has been smoothed by a
Gaussian kernel).  The optical morphology of the possible stellar
counterpart is highly irregular, appearing as a number of low surface
brightness knots of emission oriented from southwest to northeast.
Remarkably, this same morphology is apparent in the color GALEX image
presented in panel (c); the blue color is indicative of a significant
far-UV flux.  Note that AGC\,229166 is not detected with significance
in the GALEX images.

The \HI\ associated with AGC\,226178 shows properties that are akin to
those of low-mass dwarf galaxies.  The neutral hydrogen is centrally
concentrated, reaching a maximum column density of
1\,$\times$\,10$^{20}$ cm$^{-2}$.  While the source is unresolved by
these VLA observations, there is a clearly defined velocity gradient
spanning $\sim$10 \kms.  The total ALFALFA flux integral implies an
\HI\ mass of $\sim$4\,$\times$\,10$^{7}$ \msun.  Our VLA images
recover $\sim$40\% of this flux; there is minor contamination from
RFI, but it is likely that some of the \HI\ detected by Arecibo is
distributed over larger angular scales than those probed by the present
VLA observations.  

The optical and UV properties of AGC\,226178 are extracted at the
position of the \HI\ centroid and using the parameters given in
Table~\ref{table_observed}.  Note that the elliptical aperture size
closely matches the UV and optical morphology of the source.  Using
the SDSS images, we derive M$_{\rm HI}$/L$_{\rm g}$ $=$ 7; M$_{\rm
  HI}$/L$_{\rm B}$ $=$ 6, and M$_{\rm HI}$/M$_{\star}$ $=$ 30.  The
far-UV flux implies a recent SFR$_{\rm FUV}$ $=$ 0.0017
\msun\,yr$^{-1}$.

We conclude that the \HI\ source AGC\,226178 is a dwarf galaxy.  We
base this conclusion on two primary lines of evidence.  First, there
is excellent positional agreement between the \HI\ centroid and the
UV-bright stellar component.  Second, the UV morphology of the
putative stellar component suggests an inclined disk morphology.
Assuming circular rotation, the orientation of this disk is
perpendicular to the isovelocity contours shown in
Figure~\ref{images-226178}(d); this is exactly what would be expected
in a rotating disk seen nearly edge-on.  

If the \HI\ source AGC\,226178 were instead associated with the very
low surface brightness object labeled as AGC\,229166 in
Figures~\ref{images-226178-full} and \ref{images-226178}, then
significant interpretive challenges would arise.  First, the excellent
positional and kinematic agreements noted above would need to occur by
chance with an unrelated background source.  Second, the
$\sim$1\arcmin\ angular offset between the \HI\ source and the
centroid of AGC\,229166 corresponds to a minimum $\sim$5 kpc physical
separation (assuming AGC\,226178 and AGC\,229166 are at the same
distance); moving $\sim$10$^7$ \msun\ of \HI\ over such a distance
would require energies that could likely only arise during a 
gravitational interaction or ram pressure stripping event.  

An alternative but speculative interpretation is one in which
AGC\,226178 and AGC\,229166 are in fact the same physical system.
There are very weak suggestions of spiral structure in the smoothed
SDSS-r band image of AGC\,229166 shown in
Figure~\ref{images-226178}(d).  It is conceivable that this is a
nearly face-on, gas-poor system.  In such a scenario, AGC\,226178
could be an outer spiral arm that has experienced some sort of infall
or interaction episode that has ignited recent star formation.  While
such a scenario seems physically unlikely, it cannot be completely
ruled out with the present data.

\subsubsection{AGC\,233638}
\label{S3.3.2}

\begin{figure}
\epsscale{1}
\plotone{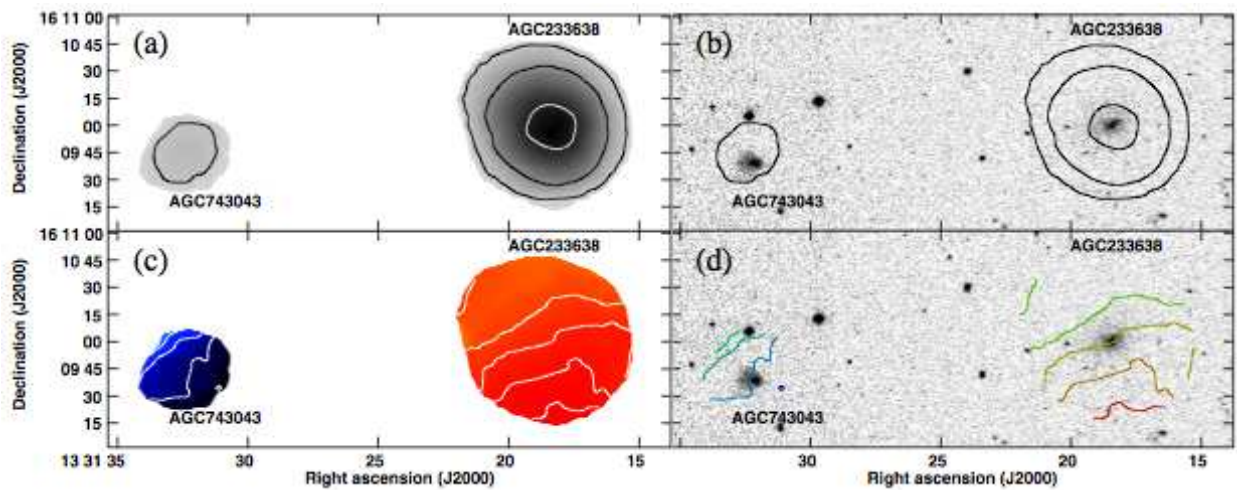}
\caption{\HI\ and SDSS R-band images of the AGC\,233638 field.  Panel
  (a) shows \HI\ column density contours at the (0.5, 1,
  2)\,$\times$\,10$^{20}$ cm$^{-2}$ levels, overlaid on a greyscale
  representation of the image.  Panel (c) shows \HI\ isovelocity
  contours at the (7330, 7335, 7340, 7345, 7400, 7405, 7410, 7415)
  \kms\ levels overlaid on a color representation of the image; the
  color bar is set to encompass the full velocity range of \HI\ gas in
  the two detected galaxies.  Panels (b) and (d) show the SDSS r-band
  image of the AGC\,226178 field, displayed in a logarithmic stretch
  to highlight low surface brightness emission; the contours in (b)
  are the same as in panel (a), while the contours in (d) are the same
  as in panel (c). The two systems in the field are labeled.  The
  \HI\ beam size is 60\arcsec; at this angular resolution,
  AGC\,743043 is unresolved, while AGC\,233638 is
  marginally resolved.}
\label{images-233638-full}
\end{figure}

AGC\,233638 was classified as a pair of galaxies with very low optical
surface brightness.  It is the most distant source in our pilot
sample.  The total ALFALFA \HI\ flux integral (of which $\sim$65\% is
recovered in our VLA observations) implies an \HI\ mass of
3.3\,$\times$\,10$^{9}$ \msun.  Figure~\ref{images-233638-full} shows
that the VLA observations resolve these two sources (AGC\,233638 and
AGC\,743043), which are separated by $\sim$3.3\arcmin\ ($\sim$108 kpc
at the adopted distance of 112 Mpc) and by less than 100 \kms.  The
SDSS-r band imaging allows a clear identification of an associated
optical counterpart with each \HI\ peak.  AGC\,233638 was included in
the ``Almost Dark'' sample due to the ambiguous nature of multiple
possible optical counterparts; the higher spatial resolutions
presented here allow us to characterize the source with confidence.

\begin{figure}
\epsscale{1}
\plotone{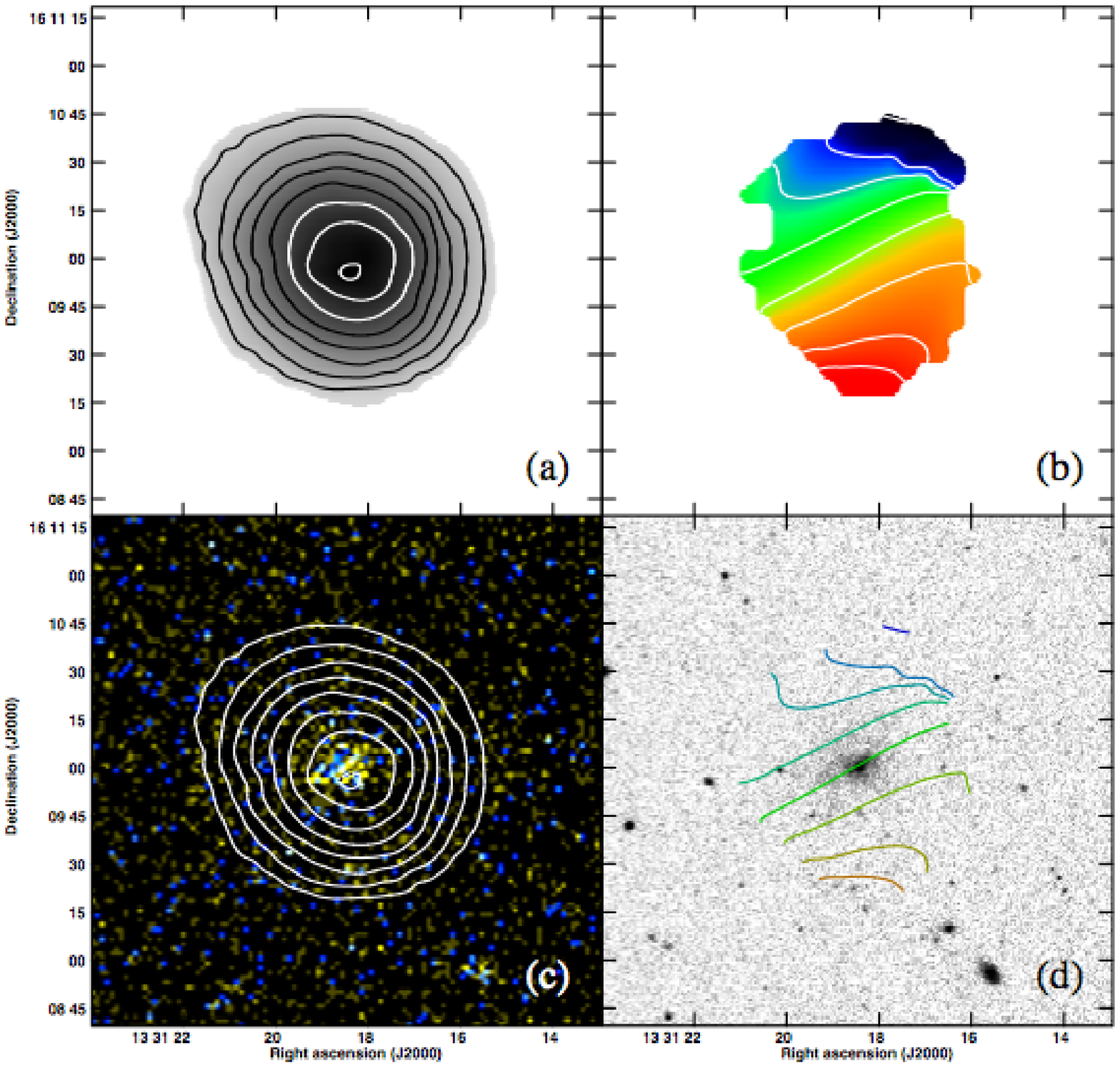}
\caption{\HI, SDSS R-band, and GALEX images of AGC\,233638.  Panel (a)
  shows \HI\ column density contours at the (0.5, 0.75, 1.0, 1.25,
  1.5, 1.75, 2.0, 2.25)\,$\times$\,10$^{20}$ cm$^{-2}$ levels,
  overlaid on a greyscale representation of the image.  Panel (b)
  shows \HI\ isovelocity contours from 7385 km\,s$^{-1}$ to 7420
  km\,s$^{-1}$ in intervals of 5 km\,s$^{-1}$, overlaid on a color
  representation of the velocity field image.  Panel (c) shows a color
  representation of the GALEX UV images, while panel (d) shows the
  SDSS r-band image (displayed in a logarithmic stretch to highlight
  low surface brightness emission); the contours in (c) are the same
  as in panel (a), while the contours in (d) are the same as in panel
  (b).  The velocity field was derived by fitting Gaussian functions
  to the full data cube.  The \HI\ beam size is 60\arcsec; AGC\,233638
  is slightly resolved at this angular resolution.}
\label{images-233638}
\end{figure}

As Figure~\ref{images-233638} shows, the candidate ``Almost Dark''
source is a star-forming dwarf irregular galaxy.  There is coherent
and well-ordered rotation spanning a projected velocity width of
$\sim$40 \kms.  The \HI\ surface density maximum is exactly co-spatial
with the GALEX centroid.  While the GALEX images are shallow, they
nonetheless allow a robust measurement of the UV fluxes in both bands
(see SDSS and GALEX photometry in Table~\ref{table_observed}).  Based
on the far-UV flux, this source has undergone significant star
formation during the most recent $\sim$100 Myr (SFR$_{\rm FUV}$ $=$
0.05 M$_{\odot}$\,yr$^{-1}$).  
The M$_{\rm HI}$/L and M$_{\rm HI}$/M$_{\star}$ ratios for AGC\,233638
are typical of those found in star-forming galaxies: 
M$_{\rm HI}$/L$_{\rm g}$ $=$ 3, M$_{\rm HI}$/L$_{\rm B}$ $=$ 3, 
M$_{\rm HI}$/M$_{\star}$ $=$ 14.  We
conclude that AGC\,233638 is a very intriguing star-forming dwarf
irregular galaxy; it has an absolute magnitude (M$_{\rm B}$ $\simeq$
$-$17) roughly equal to that of the Large Magellanic Cloud, a total
\HI\ mass (M$_{\rm HI}$ = 3.2\,$\times$\,10$^{9}$ \msun) roughly equal
to that of the Milky Way, and a stellar mass (M$_{\star}$ $=$ 2.2\,$\times$\,10$^8$
\msun) typical of those found in dwarf galaxies.

A bright nuclear point source is coincident with the optical
counterpart of AGC\,233638; an SDSS spectrum of this bright central
knot is available. The SDSS pipeline software proposes a redshift of
$z$=0.940.  Inspection of the SDSS spectrum suggests that this
high-redshift classification may be erroneous; no obvious emission or
absorption lines agree with this redshift.  Further, this would be
highly inconsistent with the present observations of the velocity of
the neutral interstellar medium.  If this compact source is not
associated with AGC\,233638 itself, then the optical luminosity and
resulting stellar mass of AGC\,233638 will both decrease, raising the
M$_{\rm HI}$/L and M$_{\rm HI}$/M$_{\star}$ ratios. Follow-up optical
spectroscopy of this compact source would be very valuable.

\section{Discussion}
\label{S4}

\subsection{The ``Almost Dark'' Sources in Context}
\label{S4.1}

The five ``Almost Dark'' sources studied in this work are
representative of the sample of ALFALFA-discovered systems that are
not uniquely identified with optical counterparts using ALFALFA
data alone.  As a result, most harbor unusually high \HI\ mass to
light ratios.  In the present work, we use multi-wavelength follow-up
observations to determine the nature of these objects: 4 sources are
likely galaxies, while one source is likely a tidal remnant.

\begin{figure}
\epsscale{1}
\plotone{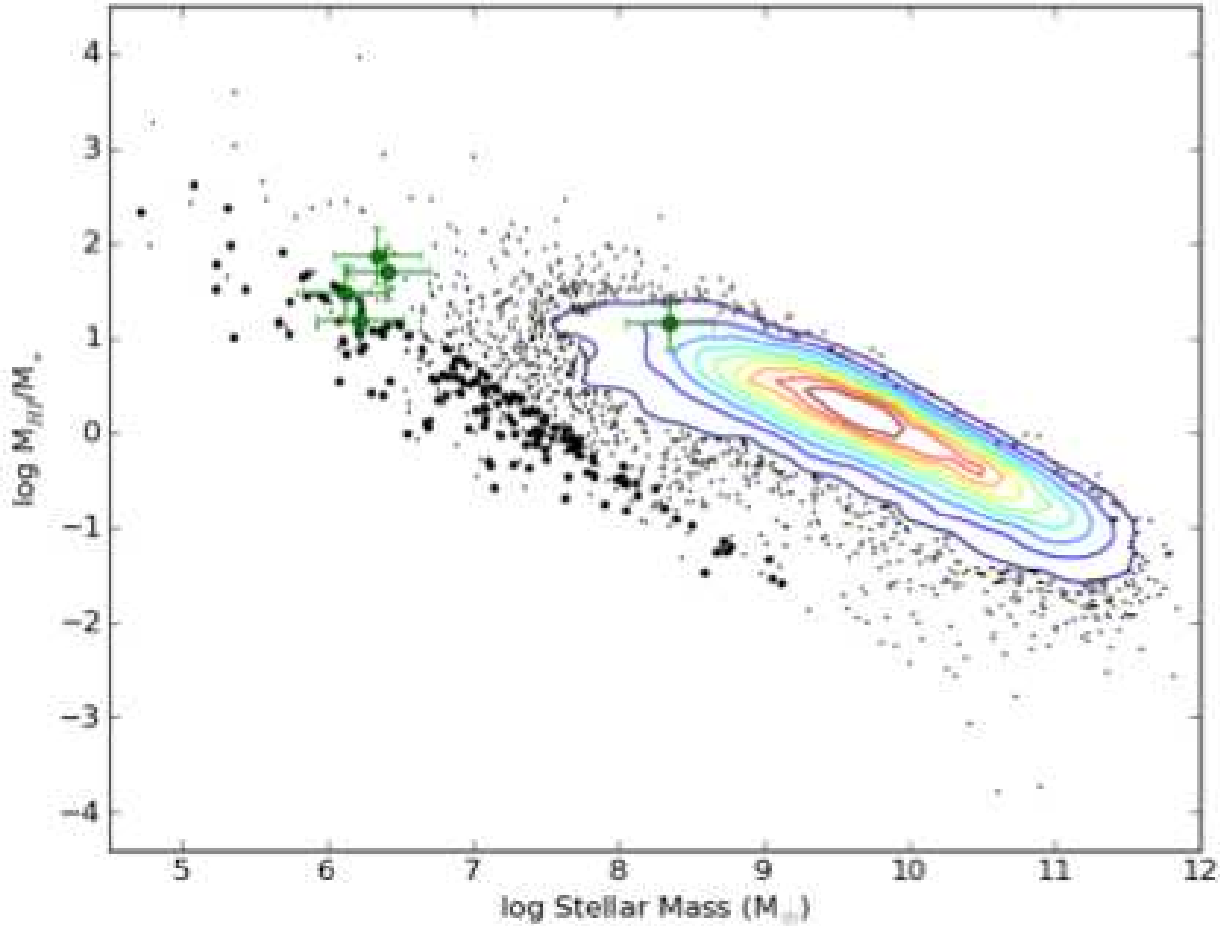}
\caption{Plot of the logarithm of the neutral hydrogen to stellar mass
  ratio, M$_{\rm HI}$/M$_{\star}$, versus the logarithm of the stellar
  mass, for all members of the $\alpha$.40 catalog (Haynes
  \etal\ 2011).  All small black points and areas within contours
  represent galaxies with reliable SDSS optical counterparts in the
  $\alpha$.40 catalog.  The contours show number densities in
  intervals of 10\%, from 10\% to 90\% of the number density maximum
  (which occurs at log(M$_{\rm HI}$) $=$ 9.8; see Haynes \etal\ 2011).
  The dwarf galaxies from the Huang \etal\ (2012) sample are shown as
  thick black points; the five ``Almost Dark'' sources studied in this
  work are shown in green.  With the exception of AGC\,233638 (which
  may have contaminated photometry; see detailed discussion in
  \S~\ref{S1} and \S~\ref{S3.3.2}), these objects preferentially
  populate the extreme upper left region of this plot.}
\label{images-a40}
\end{figure}

Keeping in mind the uncertainties that
arise because different works have measured optical properties in
different ways, the systems studied here can be compared to both the
sample of ALFALFA dwarf galaxies in \citet{huang12} and to the full
$\alpha$.40 database presented in \citet{haynes11}.  In
Figure~\ref{images-a40} we thus show the logarithm of the 
M$_{\rm HI}$/M$_{\star}$ ratio versus the logarithm of the stellar
mass for all of these systems.  There is a clear trend for higher
M$_{\rm HI}$/M$_{\star}$ ratios in systems of fainter absolute
magnitude (lower stellar mass), although the scatter is significant
(becoming more pronounced as stellar mass decreases).  With the
exception of AGC\,208399, the five sources studied here preferentially
populate the extreme upper left region of this plot; these are
gas-rich systems with low stellar masses.

A second comparison can be made against targeted studies of gas-rich,
low surface brightness dwarf galaxies.  For example, \citet{vanzee97}
computes the M$_{\rm HI}$/L ratios in a sample of seven low surface
brightness dwarfs and a comparison sample of five dwarfs with more
typical hydrogen mass to light ratios.  All twelve of these systems
have M$_{\rm HI}$/L$_{\rm B}$ values near unity (the lowest ratio
being 0.6 in UGC\,891 and the highest ratio being 6.0 in UGC\,5716).
The M$_{\rm HI}$/L$_{\rm B}$ values of the four ``Almost Dark''
systems in the present study that appear to be bona fide galaxies (see
Table~\ref{table_derived}) range from 3 (AGC\,233638) to 9
(AGC\,193953).  The galaxies in the present work thus harbor M$_{\rm
  HI}$/L$_{\rm B}$ that are comparable to the most extreme systems in
\citet{vanzee97}.

Further comparisons can made with studies of low surface brightness
galaxies that display elevated M$_{\rm HI}$/L ratios (e.g., {Longmore
  \etal\ 1982}\nocite{longmore82}; {van~Zee
  \etal\ 1995}\nocite{vanzee95}; {de~Blok
  \etal\ 1996}\nocite{deblok96}; {Matthews \&
  Gallagher 1997}\nocite{matthews97}; {O'Neil \etal\ 2000}\nocite{oneil00};
{Doyle \etal\ 2005}\nocite{doyle05}).  While some physical
characteristics of the present ``Almost Dark'' objects are similar to
those found for objects in these previous studies, the ALFALFA
``Almost Dark'' sources constitute a different class of object.  For
example, all of the low surface brightness sources in the
\citet{deblok96} sample that are within the ALFALFA footprint are
detected in \HI, and all of them are assigned an optical counterpart
by comparison with SDSS imaging.  In contrast, the ALFALFA ``Almost
Dark'' sources represent the most extreme objects ($<$0.5\% of the
full ALFALFA catalog) for which an optical counterpart remains
elusive, even when comparing the sensitive ALFALFA and SDSS/DSS2B
survey products.

\subsection{The Physical Characteristics of the ``Almost Dark'' Sources}
\label{S4.2}

It is tempting to study the nature of the recent (within $\sim$100
Myr) star formation in those systems in which star formation is
apparent based on a measurable GALEX far-UV flux (AGC\,208399,
AGC\,226178, AGC\,233638). However, with the angular resolution
afforded by these D configuration observations, all of these sources
are only marginally resolved (if at all).  The resulting \HI\ column
densities and corresponding mass surface densities will necessarily be
lower limits that will increase upon observation at higher angular
resolution.  Taken at face value, the observed column density maxima
[N$_{\rm HI}$ $=$ (1-2)\,$\times$\,10$^{20}$ cm$^{-2}$; see
  Figures~\ref{images-208399}, \ref{images-226178}, and
  \ref{images-233638}] are significantly lower than the peak column
densities observed in the more nearby galaxies of the \citet{vanzee97}
sample [N$_{\rm HI}$ $=$ (7-30)\,$\times$\,10$^{20}$ cm$^{-2}$].
Higher spatial resolution observations of these ``Almost Dark''
sources would be valuable for studying the nature of the star
formation law in the extreme M$_{\rm HI}$/L regime by allowing us to
resolve the \HI\ surface densities in the disks and correlating these
values with well-understood tracers of ongoing star formation.

The elevated M$_{\rm HI}$/L and M$_{\rm HI}$/M$_{\star}$ ratios in the
four non-tidal ``Almost Dark'' candidates raise interesting questions
about galaxy evolution.  The \HI\ masses are not unusually low for any
of these sources; as with all known non-spheroidal galaxies, an
\HI\ reservoir larger than 10$^7$ \msun\ appears to be at least
partially converted into stars.  However, in these sources, this
conversion appears to be extremely inefficient.  Using the UV-based
SFRs tabulated in Table~\ref{table_derived}, the gas consumption
timescales exceed the Hubble time for all sources.  These ``Almost
Dark'' galaxies are an extreme case of inefficient star formation,
similar to, but more dramatic than, what is found in the outer regions
of spiral disks and in gas-dominated dwarfs \citep[c.f.,][]{bigiel08}.

The angular resolution of the \HI\ observations presented here results
in physical resolution elements ranging from 2.5 kpc (AGC\,208399) to
32.6 kpc (AGC\,233638).  This precludes rotation curve work and a
detailed mass decomposition.  It is iteresting to note in this context
that the objects identified as galaxies do show irregular kinematics
and morphology compared to dynamically relaxed systems. For example,
AGC\,208399 and AGC\,226178 show lopsided kinematics; AGC\,193953
shows a misalignment between the total \HI\ intensity and the
kinematical axes; AGC\,223638 may be warped or show symmetric
(bar-like) streaming motions.  Higher resolution observations of the
nearest of these sources would facilitate an examination of the dark
and baryonic components in these systems.

\section{Conclusions and Outlook}
\label{S5}

We have presented new \HI\ spectral line observations of five sources
extracted from the ALFALFA catalog as being representative members of
the ``Almost Dark'' population: systems for which an optical
counterpart cannot be assigned using ALFALFA data alone.  These
sources harbor extreme hydrogen mass to optical luminosity ratios.
With the optical, UV, and interferometric \HI\ imaging presented in this 
work, we now have a much more clear understanding of the nature of these
objects; none of them appears to be a bona fide ``dark galaxy''.   

The \HI\ images of these sources localize the \HI\ gas and provide a
coarse representation of the rotation and dynamics of the neutral
interstellar medium.  We compare these \HI\ results with SDSS and
GALEX imaging.  In four of the five systems, a stellar counterpart is
detected that is co-spatial with the \HI\ centroid.  Each of these
objects shows signatures of coherent rotation; we classify these
systems as galaxies.

One system in this study (AGC\,208602) lacks a detectable stellar
population in SDSS and GALEX imaging.  This source has the largest
M$_{\rm HI}$/L ratio in our sample, has an irregular \HI\ morphology,
and lacks clear signs of coherent rotation.  Given its location in the
NGC\,3370 group, we conclude that this feature has a tidal origin.

The gaseous and stellar components of AGC\,226178 make it a very
intriguing system.  It is located within a projected separation of
$\sim$75 kpc from VCC\,2034 and VCC\,2037, both of which are detected
in our \HI\ imaging.  It is also in very close angular projection
($\sim$1\arcmin, or $\sim$5 kpc at the adopted distance) to
AGC\,229166, a source with very low optical surface brightness that is
not detected in \HI\ by either ALFALFA or the present observations or
in the UV in GALEX imaging.  Surprisingly, these GALEX images reveal a
significant UV flux from an irregular-morphology source that is
coincident with the \HI\ centroid of AGC\,226178.  We interpret this
source as the stellar counterpart of AGC\,226178 and thus classify it
as a dwarf galaxy.  The nature of AGC\,229166, and its possible
association with AGC\,226178, remains unclear with the present data.

The remaining three sources in this study (AGC\,193953, AGC\,208399,
and AGC\,233638) appear to be galaxies with higher than average
M$_{\rm HI}$/L ratios in comparison with those found in the samples of
\citet{vanzee97}, \citet{haynes11}, and \citet{huang12}.  Of these
sources, AGC\,193953 and AGC\,208399 are the most promising for more
detailed investigation.  With the highest M$_{\rm HI}$/L ratios of the
galaxies in this study, and very low UV-based star formation rates,
these sources can facilitate an exploration of the star formation
threshold in a sparsely-explored region of parameter space.
AGC\,233638 is a comparatively distant source with a lower mass to
light ratio; higher resolution observations of this source will be
less valuable.

As of this writing, there are roughly 50 viable ``Almost Dark''
candidate sources in the ALFALFA catalog.  The present pilot study of
five sources from this larger sample demonstrates the complexity of
the members of this heterogenous class of objects: the higher angular
resolution VLA images localize the \HI\ gas (thereby facilitating an
association with a stellar component for most of these objects) and
reveal ordered motions indicative of galaxy-like objects.  The
discovery potential of a dedicated follow-up observing campaign to
study the gaseous components of all candidate ``Almost Dark'' systems
is thus exceptionally high.

Our team is pursuing similar \HI\ observations of the full ``Almost
Dark'' sample, as well as deep optical images of the sources with
large field-of-view instruments.  For example, preliminary results
from these initiatives demonstrate significant similarities between
AGC\,226178 (studied here) and AGC\,229385 and its neighbors
(Janowiecki \etal, in preparation).  These future works hold
significant promise for furthering our understanding of the evolution
of extremely gas-rich systems.  They will allow us to explore the
nature of star formation in the high M$_{\rm HI}$/L regime; the
preliminary results from the present work suggest a very inefficient
or suppressed star formation process.  Most importantly, these
endeavors will allow us to quantify the frequency of truly ``dark''
galaxies (if they exist), using the statistically robust ALFALFA
database with well-understood sensitivity and completeness limits.

\acknowledgements
 
J.M.C. is supported by NSF grant 1211683.  J.M.C., R.A.K., C.P.M.,
N.N., and P.T. acknowledge NSF grant AST-1211005.  The ALFALFA work at
Cornell is supported by NSF grant AST-1107390 and by the Brinson
Foundation.  K.L.R. acknowledges support from NSF grant AST-0847109.
J.M.C. would like to thank the Instituto Nazionale di Astrofisica and
the Osservatorio Astronomico di Padova for their hospitality during a
productive sabbatical leave.  L.L. would like to thank Angus Wright
for technical support using his code, LAMBDAR.  G.J. started to
participate in this project as an employee of ASTRON, Dwingeloo, The
Netherlands (Nederlandse Organisatie voor Wetenschappelijk Onderzoek,
NWO).  This investigation has made use of the NASA/IPAC Extragalactic
Database (NED) which is operated by the Jet Propulsion Laboratory,
California Institute of Technology, under contract with the National
Aeronautics and Space Administration, and NASA's Astrophysics Data
System.  This investigation has made use of data from the Sloan
Digital Sky Survey (SDSS). Funding for the SDSS and SDSS-II has been
provided by the Alfred P. Sloan Foundation, the Participating
Institutions, the National Science Foundation, the U.S. Department of
Energy, the National Aeronautics and Space Administration, the
Japanese Monbukagakusho, the Max Planck Society, and the Higher
Education Funding Council for England. The SDSS Web Site is
http://www.sdss.org/.

The SDSS is managed by the Astrophysical Research Consortium for the
Participating Institutions. The Participating Institutions are the
American Museum of Natural History, Astrophysical Institute Potsdam,
University of Basel, University of Cambridge, Case Western Reserve
University, University of Chicago, Drexel University, Fermilab, the
Institute for Advanced Study, the Japan Participation Group, Johns
Hopkins University, the Joint Institute for Nuclear Astrophysics, the
Kavli Institute for Particle Astrophysics and Cosmology, the Korean
Scientist Group, the Chinese Academy of Sciences (LAMOST), Los Alamos
National Laboratory, the Max-Planck-Institute for Astronomy (MPIA),
the Max-Planck-Institute for Astrophysics (MPA), New Mexico State
University, Ohio State University, University of Pittsburgh,
University of Portsmouth, Princeton University, the United States
Naval Observatory, and the University of Washington.

\clearpage
\bibliographystyle{apj}                                                 

\begin{thebibliography}{}

\bibitem[Adams \etal(2014)]{adams14} Adams, E.A.K., Faerman, Y., Oosterloo,
T.~A., et al.\ 2014, \aap, submitted

\bibitem[Ahn \etal(2014)]{ahn14} Ahn, C.~P., Alexandroff, R., Allende
  Prieto, C., et al.\ 2014, \apjs, 211, 17

\bibitem[Bigiel \etal(2008)]{bigiel08} Bigiel, F., Leroy, A., 
Walter, F., et al.\ 2008, \aj, 136, 2846

\bibitem[Blanton \& Roweis(2007)]{blanton07} Blanton, M.~R., \&
  Roweis, S.\ 2007, \aj, 133, 734

\bibitem[Cannon \etal(2011)]{cannon11} Cannon, J.~M., Giovanelli, R.,
  Haynes, M.~P., et al.\ 2011, \apjl, 739, L22

\bibitem[Cappellari \etal(2011)]{cappellari11} Cappellari, M.,
  Emsellem, E., Krajnovi{\'c}, D., et al.\ 2011, \mnras, 413, 813

\bibitem[de Blok \etal(1996)]{deblok96} de Blok, W.~J.~G., McGaugh,
  S.~S., \& van der Hulst, J.~M.\ 1996, \mnras, 283, 18

\bibitem[Dowell(2010)]{dowell11} Dowell, J.~D.\ 2010, 
Ph.D.~Thesis, Indiana University

\bibitem[Doyle \etal(2005)]{doyle05} Doyle, M.~T., Drinkwater, M.~J.,
  Rohde, D.~J., et al.\ 2005, \mnras, 361, 34

\bibitem[Duc \& Bournaud(2008)]{duc08} Duc, P.-A., \& Bournaud,
  F.\ 2008, \apj, 673, 787

\bibitem[English \etal(2010)]{english10} English, J., Koribalski, B.,
  Bland-Hawthorn, J., Freeman, K.~C., \& McCain, C.~F.\ 2010, \aj,
  139, 102

\bibitem[Giovanelli \etal(2005)]{giovanelli05} Giovanelli, R., 
Haynes, M.~P., Kent, B.~R., et al.\ 2005, \aj, 130, 2598 

\bibitem[Haynes \etal(2007)]{haynes07} Haynes, M.~P., Giovanelli, R.,
  \& Kent, B.~R.\ 2007, \apjl, 665, L19

\bibitem[Haynes \etal(2011)]{haynes11} Haynes, M.~P., 
Giovanelli, R., Martin, A.~M., et al.\ 2011, \aj, 142, 170 

\bibitem[Huang \etal(2012)]{huang12} Huang, S., Haynes, M.~P., 
Giovanelli, R., et al.\ 2012, \aj, 143, 133

\bibitem[Jorsater \& van Moorsel(1995)]{jorsater95} Jorsater, S., \& van
    Moorsel, G.~A.\ 1995, \aj, 110, 2037 

\bibitem[Kent \etal(2008)]{kent08} Kent, B.~R., Giovanelli, 
R., Haynes, M.~P., et al.\ 2008, \aj, 136, 713 

\bibitem[Kent \etal(2009)]{kent09} Kent, B.~R., Spekkens, K.,
  Giovanelli, R., et al.\ 2009, \apj, 691, 1595

\bibitem[Kent(2010)]{kent10} Kent, B.~R.\ 2010, \apj, 725, 2333

\bibitem[Lee-Waddell \etal(2014)]{leewaddell14} Lee-Waddell, K., Spekkens, K., 
Cuillandre, J.-C., et al.\ 2014, \mnras, 443, 3601

\bibitem[Longmore \etal(1982)]{longmore82} Longmore, A.~J., Hawarden,
  T.~G., Goss, W.~M., Mebold, U., \& Webster, B.~L.\ 1982, \mnras,
  200, 325

\bibitem[Martin \etal(2005)]{martin05} Martin, D.~C., Fanson, 
J., Schiminovich, D., et al.\ 2005, \apjl, 619, L1 

\bibitem[Martin \etal(2010)]{martin10} Martin, A.~M., Papastergis, E.,
  Giovanelli, R., et al.\ 2010, \apj, 723, 1359

\bibitem[Martin \etal(2012)]{martin12} Martin, A.~M., 
Giovanelli, R., Haynes, M.~P., \& Guzzo, L.\ 2012, \apj, 750, 38

\bibitem[Masters(2005)]{masters05} Masters, K.~L.\ 2005, 
Ph.D.~Thesis, Cornell University

\bibitem[Matsuoka \etal(2012)]{matsuoka12} Matsuoka, Y., Ienaka, N.,
  Oyabu, S., Wada, K., \& Takino, S.\ 2012, \aj, 144, 159

\bibitem[Matthews \& Gallagher(1997)]{matthews97} Matthews, L.~D., \&
  Gallagher, J.~S., III 1997, \aj, 114, 1899

\bibitem[Mihos \etal(2012)]{mihos12} Mihos, J.~C., Keating, K.~M.,
  Holley-Bockelmann, K., Pisano, D.~J., \& Kassim, N.~E.\ 2012, \apj,
  761, 186

\bibitem[Minchin \etal(2005)]{minchin05} Minchin, R., Davies, J.,
  Disney, M., et al.\ 2005, \apjl, 622, L21

\bibitem[Minchin \etal(2007)]{minchin07} Minchin, R., Davies, J.,
  Disney, M., et al.\ 2007, \apj, 670, 1056

\bibitem[O'Neil \etal.(2000)]{oneil00} O'Neil, K., Bothun, G.~D., \&
  Schombert, J.\ 2000, \aj, 119, 136

\bibitem[Oosterloo \etal(2013)]{oosterloo13} Oosterloo, T.~A., Heald,
  G.~H., \& de Blok, W.~J.~G.\ 2013, \aap, 555, L7

\bibitem[Papastergis \etal(2011)]{papastergis11} Papastergis, E.,
  Martin, A.~M., Giovanelli, R., \& Haynes, M.~P.\ 2011, \apj, 739, 38

\bibitem[Salim \etal(2007)]{salim07} Salim, S., Rich, R.~M., Charlot,
  S., et al.\ 2007, \apjs, 173, 267

\bibitem[Schlafly \& Finkbeiner(2011)]{schlafly11} Schlafly, E.~F., \&
  Finkbeiner, D.~P.\ 2011, \apj, 737, 103

\bibitem[Serra \etal(2013)]{serra13} Serra, P., Koribalski, B., Duc,
  P.-A., et al.\ 2013, \mnras, 428, 370

\bibitem[van Zee \etal(1995)]{vanzee95} van Zee, L., Haynes, M.~P., \&
  Giovanelli, R.\ 1995, \aj, 109, 990

\bibitem[van Zee \etal(1997)]{vanzee97} van Zee, L., Haynes, M.~P.,
  Salzer, J.~J., \& Broeils, A.~H.\ 1997, \aj, 113, 1618

\bibitem[Walter \etal(2005)]{walter05} Walter, F., Skillman, E.~D., \&
  Brinks, E.\ 2005, \apjl, 627, L105

\bibitem[Walter \etal(2008)]{walter08} Walter, F., Brinks, E., de
  Blok, W.~J.~G., et al.\ 2008, \aj, 136, 2563

\bibitem[Zibetti \etal(2009)]{zibetti09} Zibetti, S., Charlot, S., \&
  Rix, H.-W.\ 2009, \mnras, 400, 1181

\end{thebibliography}



\clearpage
\begin{deluxetable}{lcccccccc} 
\tablecaption{Basic and Global HI Properties of the Candidate ``Almost Dark'' Systems} 
\tablewidth{0pt}  
\tablehead{ 
\colhead{AGC}    &\colhead{RA\tablenotemark{a}}      &\colhead{Dec\tablenotemark{a}} &\colhead{Offset\tablenotemark{b}}     &\colhead{S$_{\rm HI}$\tablenotemark{c}} &\colhead{V$_{\rm sys}$} &\colhead{W$_{\rm 50}$} &\colhead{D\tablenotemark{d}} &\colhead{log(M$_{\rm HI}$)}\\
\colhead{Number} &\colhead{(J2000)} &\colhead{(J2000)} &\colhead{(arcmin)}  &\colhead{(Jy\,\kms)}         &\colhead{(\kms)}       &\colhead{(\kms)}      &\colhead{(Mpc)}   &\colhead{(\msun)}}   
\startdata      
193953\tablenotemark{e}           & 143.6444         & 12.394      & 1.32     & 0.43                        & 2591                  &32                 &39.9 & 8.21 \\
208399\tablenotemark{e}           & 160.0417         &  4.910      & 0.36     & 1.06                        &  747                  &31                 &10.1 & 7.41 \\
208602\tablenotemark{f}           & 163.6558         & 17.635      & 0.92     & 1.45                        & 1093                  &50                 &18.5  & 8.07 \\
226178\tablenotemark{e,g}         & 191.6782         & 10.369      & 1.33     & 0.62                        & 1581                  &28                 &16.5 & 7.60 \\
233638\tablenotemark{e}           & 202.8267         & 16.167      & 0.05     & 1.10                        & 7587                  &62                 &112  & 9.51 \\
\enddata     
\label{table_basic}
\tablenotetext{a}{RA and Dec of the \HI\ centroid measured in the VLA images.}
\tablenotetext{b}{Angular offsets between ALFALFA centroid coordinates and VLA centroid coordinates.}
\tablenotetext{c}{Total integrated \HI\ line flux density measured by ALFALFA in Jy \kms.}
\tablenotetext{d}{Distances based on the local flow model by \citet{masters05} and \citet{martin10}.}
\tablenotetext{e}{Very low surface brightness optical counterpart.}
\tablenotetext{f}{Positional offset from early-type galaxy NGC\,3457.}
\tablenotetext{g}{Multiple possible optical counterparts.}
\end{deluxetable} 

\clearpage
\begin{deluxetable}{lcccccc}  
\tablecaption{Observed Properties of the Candidate ``Almost Dark'' Systems} 
\tablewidth{0pt}  
\tablehead{ 
\colhead{AGC}    &\colhead{Aperture\tablenotemark{a}} &\colhead{m$_{\rm FUV}$\tablenotemark{b}} &\colhead{m$_{\rm NUV}$\tablenotemark{c}} &\colhead{m$_{\rm g,SDSS}$\tablenotemark{d}} &\colhead{m$_{\rm g}$\tablenotemark{e}} &\colhead{m$_{\rm r}$\tablenotemark{f}}\\
\colhead{Number} &\colhead{Size}                      &\colhead{(mag.)} &\colhead{(mag.)} &\colhead{(mag.)} &\colhead{(mag.)} &\colhead{(mag.)}}   
\startdata      
193953  &  16"$\times$11" &  \ N/A          & 20.0$\pm$0.3 & 20.8$\pm$0.1   & 20.2$\pm$0.3 &  20.1$\pm$0.3 \\
208399  &  26"$\times$18" &  \ 21.6$\pm$0.3 & 19.9$\pm$0.3 & 20.00$\pm$0.03 & 18.6$\pm$0.2 &  18.3$\pm$0.2 \\

208602  &  15"$\times$15" &  $>$21.4          & 20.3$\pm$0.3 &     ---        &  $>$20.5       &  $>$21.0        \\
226178  &  42"$\times$8"  &  \ 19.6$\pm$0.2 & 20.2$\pm$0.3 & 22.4$\pm$0.1   & 19.4$\pm$0.3 &  19.2$\pm$0.3 \\
233638  &  25"$\times$18" &  \ 19.9$\pm$0.2 & 19.3$\pm$0.2 & 18.72$\pm$0.01 & 18.0$\pm$0.2 &  17.8$\pm$0.2 \\
\enddata     
\label{table_observed}
\tablenotetext{a}{Semi-major and semi-minor axis lengths of photometric aperture.}
\tablenotetext{b}{Apparent integrated magnitude of the source in the GALEX FUV band.  No Galactic absorption correction has been applied.}
\tablenotetext{c}{Apparent integrated magnitude of the source in the GALEX NUV band.  No Galactic absorption correction has been applied.}
\tablenotetext{d}{SDSS catalog model magnitude.}
\tablenotetext{e}{Apparent integrated magnitude of the source in the SDSS g band.  No Galactic absorption correction has been applied.}
\tablenotetext{f}{Apparent integrated magnitude of the source in the SDSS r band.  No Galactic absorption correction has been applied.}
\end{deluxetable}   

\clearpage
\begin{deluxetable}{lccccccc}  
\tablecaption{Derived Properties of the Candidate ``Almost Dark'' Systems} 
\tablewidth{0pt}  
\tablehead{ 
\colhead{AGC} &\colhead{L$_{\rm g}$\tablenotemark{a}} &\colhead{L$_{\rm B}$\tablenotemark{b}} &\colhead{M$_{\star}$\tablenotemark{c}} &\colhead{M$_{\rm HI}$/L$_{\rm g}$\tablenotemark{d}} &\colhead{M$_{\rm HI}$/L$_{\rm B}$\tablenotemark{e}} &\colhead{M$_{\rm HI}$/M$_{\star}$\tablenotemark{f}} &\colhead{SFR$_{\rm FUV}$\tablenotemark{g}}\\
\colhead{Number} &\colhead{(L$_{\odot}$)} &\colhead{(L$_{\odot}$)}  &\colhead{(M$_{\odot}$)} &\colhead{(M$_{\odot}$/L$_{\odot}$)}  &\colhead{(M$_{\odot}$/L$_{\odot}$)} &\colhead{(M$_{\odot}$/M$_{\odot}$)} &\colhead{(M$_{\odot}$\,yr$^{-1}$)}}  
\startdata      
193953  &   1.6\,$\times$\,10$^{7}$ &   1.9\,$\times$\,10$^{7}$ &    2.2\,$\times$\,10$^{6}$ &    10  &      9   &    75 &    $\ldots$\\
208399  &   4.4\,$\times$\,10$^{6}$ &   5.0\,$\times$\,10$^{6}$ &    1.7\,$\times$\,10$^{6}$ &     6  &      5   &    15 &    5.1x10$^{-5}$\\
208602  &$<$2.6\,$\times$\,10$^{6}$ &$<$2.8\,$\times$\,10$^{6}$ & $<$2.5\,$\times$\,10$^{6}$ & $>$48  &  $>$45   & $>$49 & $<$2.6x10$^{-4}$ \\
226178  &   5.7\,$\times$\,10$^{6}$ &   6.7\,$\times$\,10$^{6}$ &    1.3\,$\times$\,10$^{6}$ &     7  &      6   &    30 &    0.0017\\
233638  &   9.6\,$\times$\,10$^{8}$ &   1.1\,$\times$\,10$^{9}$ &    2.2\,$\times$\,10$^{8}$ &     3  &      3   &    14 &    0.047\\
\enddata     
\label{table_derived}
\tablenotetext{a}{g-band luminosity, corrected for Galactic extinction via \citet{schlafly11}, derived from g band magnitudes extracted from downloaded SDSS images using LAMBDAR.}
\tablenotetext{b}{B-band luminosity, corrected for Galactic extinction via \citet{schlafly11}, derived from g and r band magnitudes extracted from downloaded SDSS images using LAMBDAR.}
\tablenotetext{c}{Stellar mass, derived using the LAMBDAR r-band magnitudes, g-r colors, and the formalism in \citet{zibetti09}.}
\tablenotetext{d}{Ratio of \HI\ mass in \msun\ to SDSS g-band luminosity in L$_{\odot}$.}
\tablenotetext{e}{Ratio of \HI\ mass in \msun\ to B-band luminosity in L$_{\odot}$.}
\tablenotetext{f}{Ratio of \HI\ mass in \msun\ to stellar mass in \msun.}
\tablenotetext{g}{Far-UV star formation rate.}
\end{deluxetable}   

\clearpage

\clearpage

\clearpage

\clearpage
\begin{figure}
\epsscale{1}
\plotone{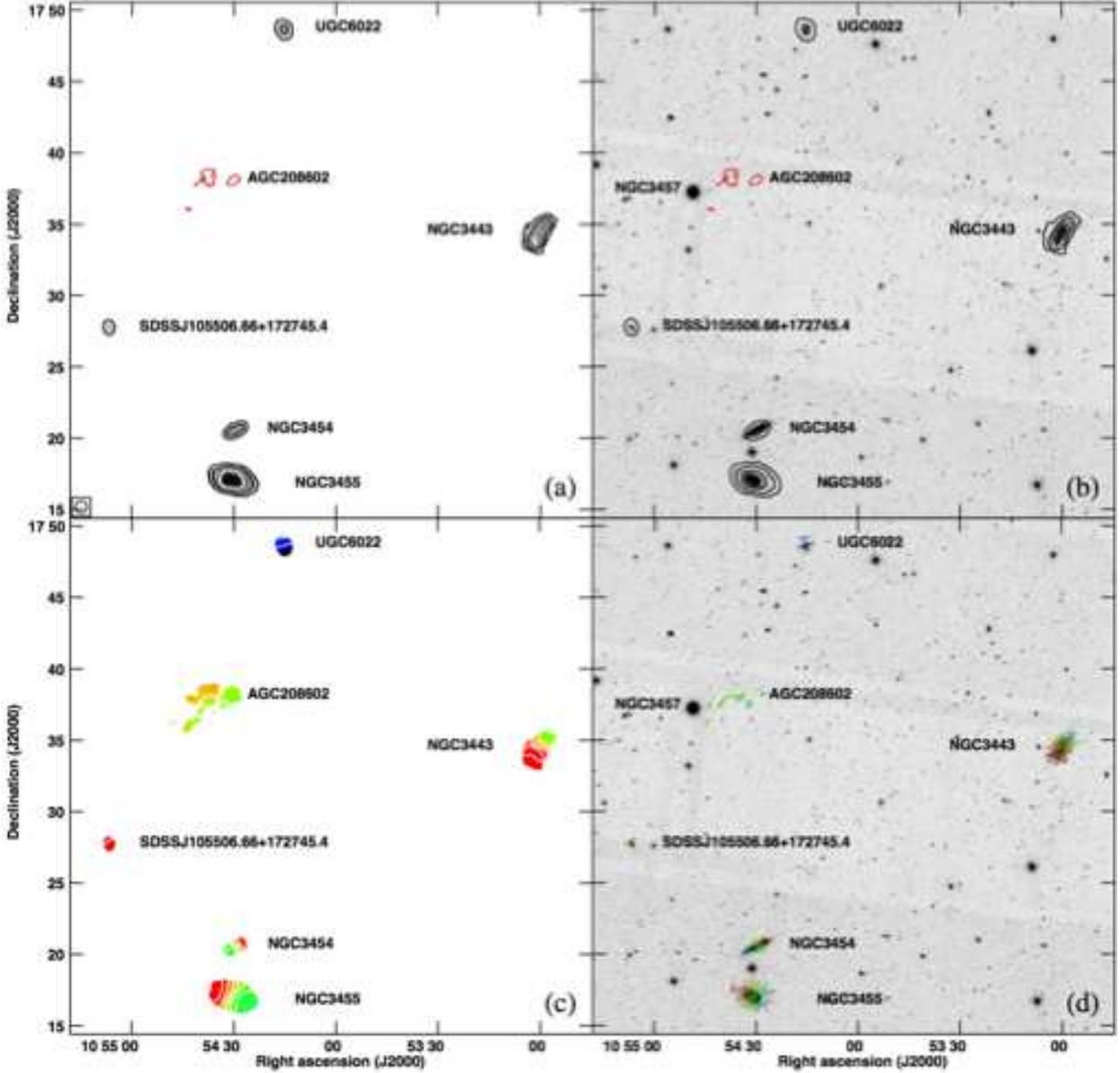}
\caption{\HI\ and SDSS R-band images of the AGC\,208602 field.  Panel
  (a) shows \HI\ column density contours overlaid at the
  3.5\,$\times$\,10$^{19}$ cm$^{-2}$ level (red, highlighting the low
  surface brightness nature of AGC\,208602) and at the (2.5, 5,
  10)\,$\times$\,10$^{20}$ cm$^{-2}$ levels (black, highlighting
  nearby sources in the field), overlaid on a greyscale representation
  of the image.  Panel (c) shows \HI\ isovelocity contours from 985
  km\,s$^{-1}$ to 1210 km\,s$^{-1}$ in intervals of 35 km\,s$^{-1}$,
  overlaid on a color representation of the velocity field image; the
  color bar is set to encompass the full velocity range of \HI\ gas in
  the six detected sources.  Panels (b) and (d) show the SDSS r-band
  image of the AGC\,208602 field, displayed in a logarithmic stretch
  to highlight low surface brightness emission; the contours in (b)
  are the same as in panel (a), while the contours in (d) are the same
  as in panel (c).  The six gas-bearing systems in the field are
  labeled, as is the early-type galaxy NGC3457.  The \HI\ beam size is
  49\arcsec, as shown in the bottom left of panel (a).}
\label{images-208602-full}
\end{figure}

\clearpage

\clearpage

\clearpage

\clearpage

\clearpage

\clearpage

\end{document}